\documentclass[reprint,amsmath,amssymb,aps,prb,superscriptaddress]{revtex4-1}
\pdfoutput=1
\usepackage{array}
\usepackage{bm}
\usepackage{graphicx}
\usepackage{hyperref}
\hypersetup{colorlinks=true,allcolors=blue}
\usepackage{natbib}
\usepackage{newtxtext}
\usepackage{newtxmath}

\begin{document}

\title{Fermionic Spin Liquid Analysis of the Paramagnetic State in Volborthite}

\author{Li Ern Chern}
\affiliation{Department of Physics, University of Toronto, Toronto, Ontario M5S 1A7, Canada}

\author{Robert Schaffer}
\affiliation{Department of Physics, University of Toronto, Toronto, Ontario M5S 1A7, Canada}

\author{Sopheak Sorn}
\affiliation{Department of Physics, University of Toronto, Toronto, Ontario M5S 1A7, Canada}

\author{Yong Baek Kim}
\affiliation{Department of Physics, University of Toronto, Toronto, Ontario M5S 1A7, Canada}
\affiliation{Canadian Institute for Advanced Research/Quantum Materials Program, Toronto, Ontario M5G 1Z8, Canada}
\affiliation{School of Physics, Korea Institute for Advanced Study, Seoul 130-722, Korea}

\begin{abstract}
Recently, thermal Hall effect has been observed in the paramagnetic state of Volborthite, which consists of distorted Kagome layers with $S=1/2$ local moments. Despite the appearance of a magnetic order below $1 \, \mathrm{K}$, the response to external magnetic field and unusual properties of the paramagnetic state above $1 \, \mathrm{K}$ suggest possible realization of exotic quantum phases. Motivated by these discoveries, we investigate possible spin liquid
phases with fermionic spinon excitations in a non-symmorphic version of the Kagome lattice, which belongs to the two-dimensional crystallographic group $p2gg$. This non-symmorphic structure is consistent with the spin model obtained in the density functional theory (DFT) calculation. Using projective symmetry group (PSG) analysis and fermionic parton mean field theory, we identify twelve distinct $\mathbb{Z}_2$ spin liquid states, four of which are found to have correspondence in the eight Schwinger boson spin liquid states we classified earlier. We focus on the four fermionic states with bosonic counterpart and find that the spectrum of their corresponding root $U(1)$ states feature spinon Fermi surfaces. The existence of spinon Fermi surface in candidate spin liquid states may offer a possible explanation of the finite thermal Hall conductivity observed in Volborthite.
\end{abstract}


\maketitle

\tableofcontents


\section{\label{introduction}Introduction}
There has been tremendous effort to understand and detect signatures of quantum spin liquid phases in frustrated magnets. The most studied example is the Heisenberg model on the isotropic Kagome lattice, which may be approximately realized in Herbertsmithite, $\mathrm{Zn Cu_3 (OH)_6 Cl_2}$, with small additional exchange interactions present. While a number of experiments on the Herbertsmithite show possible signatures of a spin liquid ground state,\cite{PhysRevLett.98.077204,PhysRevLett.98.107204,nature11659} definite confirmation still remains elusive. It is also important to demonstrate that the appearance of a spin liquid state is generic in this class of frustrated magnets and not confined to the perfectly isotropic Kagome lattice, as small distortions of the lattice or deviation from the Heisenberg model would naturally occur in many related materials. Hence it is useful to examine different materials with related lattice and magnetic structures.

Volborthite, $\mathrm{Cu_3 V_2 O_7 (OH)_2 \cdot 2 H_2 O}$, is a prominent example that hosts a distorted version of the Kagome lattice.\cite{ncomms1875} Magnetic order arises below $1 \, \mathrm{K}$, which is highly suppressed compared to the Curie-Weiss temperature of $140 \, \mathrm{K}$. The phase diagram in an external magnetic field is highly complex, and the nature of the magnetic order below $1 \, \mathrm{K}$ has not yet been fully understood.\cite{JPSJ.78.043704,PhysRevLett.103.077207,PhysRevB.83.180407,PhysRevLett.114.227202,1602.04028} Recently, the thermal Hall conductivity in Volborthite has been measured and shown to be finite in the paramagnetic state just above $1 \, \mathrm{K}$.\cite{1608.00444} This is highly unusual as Volborthite is an insulator and any heat-carrying object should be a charge-neutral excitation, which would not experience the Lorentz force via external magnetic field. Such an anomalous response may be explained if the underlying ground state is a quantum spin liquid that supports fermionic spinons with a Fermi surface. It was proposed earlier that an emergent gauge field in such spin liquid phases induces 
a fictitious magnetic field that couples to the charge-neutral spinons, which will eventually lead to a finite thermal Hall conductivity.\cite{PhysRevLett.104.066403}

In this work, motivated by the appearance of the finite thermal Hall conductivity, we investigate possible spin liquid phases with fermionic spinons in a distorted Kagome lattice appropriate for Volborthite. A microscopic spin model of Volborthite was obtained earlier from density functional theory (DFT) calculation,\cite{PhysRevLett.117.037206} which suggests that the Kagome layer can be thought of as frustrated $J_1 - J_2$ spin chains running parallel to each other and interacting through two different interchain couplings $J$ and $J'$. Analyzing the spatial symmetries of the distorted Kagome lattice described by the $J_1 - J_2 - J - J'$ model further reveals that it belongs to the non-symmorphic two-dimensional crystallographic group $p2gg$. Non-symmorphic symmetries are interesting from the theoretical perspective because they can prevent the formation of trivial band insulator even though there is an even number of electrons per unit cell. The electron fillings at which a trivial band insulator is possible for each of the $230$ space groups are tabulated in Ref.~\onlinecite{PhysRevLett.117.096404}. This is $4n$ for $p2gg$, but there are $6$ electrons per unit cell in the Volborthite Kagome layer, which is not a multiple of $4$. Therefore, the ground state of Volborthite should support either gapless excitations or topological order, by the Hastings-Oshikawa-Lieb-Schultz-Mattis (HOLSM) theorem.\cite{nphys2600} This suggests that the paramagnetic state of Volborthite cannot be a trivial state, and quantum spin liquid states with a finite excitation gap (topological order) or gapless spinon excitations are possible candidate ground states.

Previously, we studied possible quantum spin liquid phases with bosonic spinons\cite{PhysRevB.45.12377,PhysRevB.74.174423,PhysRevB.87.125127} and the magnetic orders related to them in Ref.~\onlinecite{1702.04360}. Energetic consideration suggests that a $(q,0)$ spiral order or a $(\pi,\pi)$ spin density wave is likely to be the magnetic order observed below $1 \, \mathrm{K}$ and can be obtained via the condensation of bosonic spinons in the spin liquid phases. In the current work, we use a different approach, fermionic parton mean field theory,\cite{PhysRevB.65.165113,PhysRevB.83.224413,1610.06191,PhysRevB.80.064410} to explore both gapped and gapless spin liquid states, in contrast to the bosonic theory, where only gapped spin liquids are stable. Here, we classify the fermionic spin liquid states in the non-symmorphic Kagome lattice through projective symmetry group (PSG) analysis.\cite{PhysRevB.65.165113,PhysRevB.83.224413,1610.06191} We are particularly interested in finding out which fermionic states are connected to the bosonic states that give rise to the magnetic orders mentioned above.

We identify twelve distinct $\mathbb{Z}_2$ spin liquid states that respect the space group of non-symmorphic Kagome lattice and time reversal symmetry, which we label by a number $1,\ldots,6$ followed by a letter $a$ or $b$. In addition, due to the presence of non-symmorphic symmetry, it is shown that only translationally invariant mean field states are allowed. We then explore the relation \cite{1403.0575,PhysRevB.94.035160,1603.03041,1605.05322} between the twelve fermionic states and the eight bosonic states classified in Ref.~\onlinecite{1702.04360} using the idea of symmetry fractionalization \cite{PhysRevB.87.104406,PhysRevB.90.121102} and fusion rules. Solving the vison PSG,\cite{PhysRevB.84.094419,PhysRevB.92.205131} we find that the four fermionic states 3a, 3b, 4a and 4b have bosonic correspondence, i.e. each of them can be connected to one of the eight bosonic states. In particular, one of them (4b in our notation) is connected to the most energetically favorable bosonic spin liquid state which undergoes a phase transition to $(q,0)$ spiral order upon spinon condensation. None of the fermionic states is connected to another highly competing bosonic state, from which the $(\pi,\pi)$ spin density wave arises.

We narrow down our investigation to the four fermionic states with bosonic counterparts and study their generic spectrum. We find that the spinon spectrum in the $\mathbb{Z}_2$ spin liquid of 3a, 4a and 4b have Dirac point(s) while 3b is gapped. Given the discovery of thermal Hall effect in Volborthite, we also examine the corresponding root $U(1)$ state of 3a, 3b, 4a and 4b (where the pairing amplitudes in the $\mathbb{Z}_2$ spin liquid state are turned off). This is because thermal Hall effect can naturally occur in $U(1)$ spin liquids with spinon Fermi surfaces, where the external magnetic field couples to the $U(1)$ gauge field, which in turn exerts a Lorentz force on the spinons and produces a finite transverse thermal conductivity.\cite{PhysRevLett.104.066403} We find that the root $U(1)$ spectrum of all four states features a spinon Fermi surface, which is consistent with the interpretation of the thermal conductivity experiment. Taken together, these root $U(1)$ spin liquid states, especially the 4b state related to the $(q,0)$ spiral order, may be promising candidates for the paramagnetic phase of Volborthite above $1 \, \mathrm{K}$. 

The rest of this paper is organized as follows. In Section \ref{latticehamiltonian}, we introduce the microscopic spin model and the lattice structure of Volborthite. In Section \ref{abrikosovfermionmeanfieldtheory}, we outline the fermionic parton mean field theory. In Section \ref{fermionicpsganalysis}, we review the essentials of PSG analysis and use it to search for possible fermionic spin liquid states and construct the mean field ansatzes. In Section \ref{bosonfermionmap}, we establish the mapping between bosonic and fermionic spin liquid states by solving the vison PSG. In Section \ref{result}, we analyze the generic mean field spectrum of the fermionic states that have bosonic counterparts. In Section \ref{summary}, we discuss the relevance of our results to the thermal conductivity experiment on Volborthite.

\section{\label{latticehamiltonian} Lattice and Hamiltonian}
\begin{figure}
\includegraphics[scale=0.3]{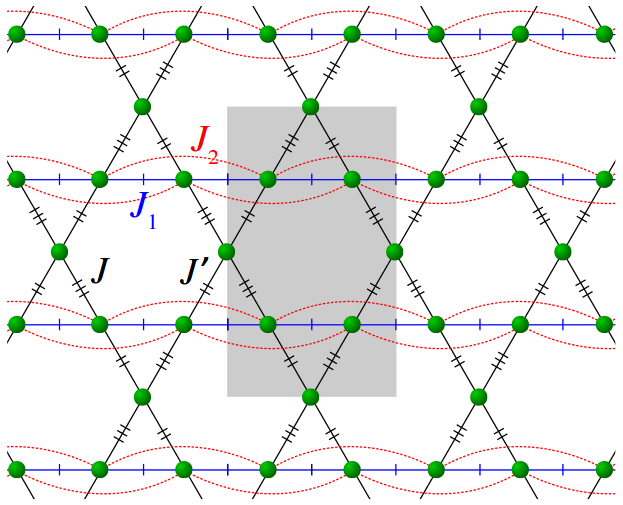}
\caption{\label{couplingfigure} The microscopic spin model of Volborthite obtained by DFT calculation,\cite{PhysRevLett.117.037206} which is the Heisenberg model \eqref{heisenberggeneral} with four leading exchange interactions $J:J':J_1:J_2=1:-0.2:-0.5:0.2$. This defines a unit cell (shaded region) with six sublattices (filled circles).}
\end{figure}

The crystal structure of Volborthite can be described by two distorted Kagome layers consisting of $\mathrm{CuO_6}$ octahedra, separated by non-magnetic $\mathrm{V_2 O_7}$ pillars and water molecules.\cite{ncomms1875} We thus analyze a single Kagome net, which has a $S=1/2$ local moment of $\mathrm{Cu}^{2+}$ ion at each site. A density functional theory (DFT) calculation \cite{PhysRevLett.117.037206} suggests that the Volborthite Kagome lattice can be viewed as frustrated $J_1 - J_2$ spin chains with two different interchain couplings $J$ and $J'$, as shown in FIG. \ref{couplingfigure}. The relative magnitude of the exchange interactions is given by $J:J':J_1:J_2=1:-0.2:-0.5:0.2$, where negative (positive) sign indicates ferromagnetic (antiferromagnetic) interaction. The Heisenberg interaction between $S=1/2$ local moments is given by
\begin{equation} \label{heisenberggeneral}
H = \sum_{ij} J_{ij} \mathbf{S}_i \cdot \mathbf{S}_j
\end{equation}
This microscopic spin model, which is dubbed the $J_1 - J_2 - J - J'$ model, defines a rectangular unit cell with six sublattices. The coordinate of any site can be written as $(x,y,s)$, where $x,y \in \mathbb{Z}$ label the unit cell and $s=1,\ldots,6$ indexes the sublattice. The chain sites have $s=1,2,4,5$, while the interstitial sites have $s=3,6$. Let us now inspect the spatial symmetries of the system. The most obvious ones are lattice translation in two independent directions, $T_x$ and $T_y$. In addition, a rotation by $\pi$, or $C_2$, also leaves the system invariant. Unlike the isotropic Kagome lattice, reflection symmetry is absent. However, a careful inspection reveals that there is a non-symmorphic glide symmetry $h$, which is a combination of reflection and half lattice translation. We henceforth refer to the Volborthite Kagome layer as non-symmorphic Kagome lattice, since it belongs to the non-symmorphic plane crystallographic group $p2gg$.

\begin{figure}
\includegraphics[scale=0.3]{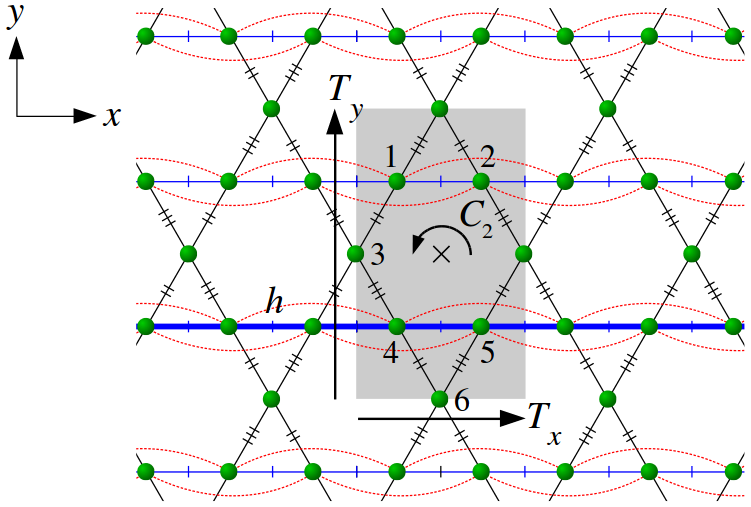}
\caption{\label{spatialsymmetry} Spatial symmetries of the non-symmorphic Kagome lattice, which are lattice translations $T_x$ and $T_y$, $\pi$-rotation $C_2$, and glide $h$. The center of rotation is marked by a cross. The action of $h$ is not shown explicitly, but the glide axis is indicated by the thick horizontal line.}
\end{figure}

There exist (infinitely many) different choices of rotation center and glide axis. We fix the rotation center at the center of the $(0,0)$ unit cell. On the other hand, we choose as the glide axis the horizontal line connecting the sites $(x,0,4)$ and $(x,0,5)$ for all $x$. The entire space group of the non-symmorphic Kagome lattice can be generated by rotation $C_2$ and glide $h$, as the lattice translations $T_x \equiv h^2$ and $T_y \equiv (C_2 h)^2$ are not independent. We nevertheless consider four spatial symmetries $T_x,T_y,C_2,h$, which are shown in FIG. \ref{spatialsymmetry}, for convenience of the subsequent analysis. We also show how a generic site $(x,y,s)$ changes under these symmetry transformations in Appendix \ref{spacegroupalgebra}.

\section{\label{abrikosovfermionmeanfieldtheory} Fermionic Parton Mean Field Theory}
In this section, we outline the fermionic parton mean field approach \cite{PhysRevB.65.165113,PhysRevB.83.224413,1610.06191,PhysRevB.80.064410} for the generic Heisenberg Hamiltonian \eqref{heisenberggeneral}. We define creation (annihilation) operators $f_{i\alpha}^{\dagger}$ ($f_{i \alpha}$) that create (annihilate) fermionic spinons with spin $\alpha \in \lbrace \uparrow , \downarrow \rbrace$ at site $i$. Then, we write the spin operator at site $i$ as
\begin{equation} \label{spinoperator}
\mathbf{S}_i = \frac{1}{2} \sum_{\alpha \beta} f_{i \alpha}^\dagger \bm{\sigma}_{\alpha \beta} f_{i \beta} ,
\end{equation}
where $\bm{\sigma}=(\sigma^x,\sigma^y,\sigma^z)$ is the vector of Pauli matrices. The fermionic operators obey the anticommutation relations
\begin{equation}
\begin{aligned} \label{anticommutation}
\left\lbrace f_{i \alpha}, f_{j \beta}^\dagger \right\rbrace &= \delta_{ij} \delta_{\alpha \beta} , \\
\left\lbrace f_{i \alpha}, f_{j \beta} \right\rbrace &= 0 = \left\lbrace f_{i \alpha}^\dagger, f_{j \beta}^\dagger \right\rbrace .
\end{aligned}
\end{equation}
Representing the spin operator as \eqref{spinoperator} in \eqref{heisenberggeneral} enlarges the original Hilbert space.\cite{PhysRevB.65.165113,PhysRevB.83.224413} To obtain the physical spin state, we have to impose the following single occupancy constraint (i.e. one fermion per site),
\begin{subequations}
\begin{align}
\sum_{\alpha} f_{i \alpha}^\dagger f_{i \alpha} &= 1 , \label{numberconstraintfermion1} \\
\sum_{\alpha \beta} f_{i \alpha} \epsilon_{\alpha \beta} f_{i \beta} &= 0 , \label{numberconstraintfermion2}
\end{align}
\end{subequations}
where $\epsilon_{\alpha \beta}$ is the antisymmetric tensor. At the mean field level, the constraint is replaced by its ground state expectation value \cite{PhysRevB.65.165113}
\begin{subequations}
\begin{align}
\sum_{\alpha} \left\langle f_{i \alpha}^\dagger f_{i \alpha} \right\rangle &= 1 , \label{numberconstraintaveragefermion1} \\
\sum_{\alpha \beta} \left\langle f_{i \alpha} \epsilon_{\alpha \beta} f_{i \beta} \right\rangle &= 0 . \label{numberconstraintaveragefermion2}
\end{align}
\end{subequations}

Next, we define the bond operators
\begin{subequations}
\begin{align}
\hat{\chi}_{ij} &= \sum_{\alpha} f_{i \alpha}^\dagger f_{j \alpha} , \label{singlethoppingfermion} \\
\hat{\Delta}_{ij} &= \sum_{\alpha \beta} f_{i \alpha} [i \sigma^y]_{\alpha \beta} f_{j \beta} , \label{singletpairingfermion} \\
\hat{E}^a_{ij} &= \sum_{\alpha \beta} f_{i \alpha}^\dagger [i \sigma^a]_{\alpha \beta} f_{j \beta} , \label{triplethoppingfermion} \\
\hat{D}^a_{ij} &= \sum_{\alpha \beta} f_{i \alpha} [i \sigma^y \sigma^a]_{\alpha \beta} f_{j \beta} , \label{tripletpairingfermion}
\end{align}
\end{subequations}
where $a=x,y,z$. $\hat{\chi}_{ij}$ ($\hat{\Delta}_{ij}$) is known as singlet hopping (pairing) channel, while $\hat{\mathbf{E}}_{ij}$ ($\hat{\mathbf{D}}_{ij}$) is known as triplet hopping (pairing) channel. The spin product can be expressed solely in terms of singlet or triplet channel as \cite{PhysRevB.80.064410}
\begin{subequations}
\begin{align}
\mathbf{S}_i \cdot \mathbf{S}_j &= -\frac{1}{4} \left( \hat{\chi}_{ij}^\dagger \hat{\chi}_{ij} + \hat{\Delta}_{ij}^\dagger \hat{\Delta}_{ij} - 1 \right) \label{spinproductsingletfermion} \\
&= \frac{1}{4} \left( \hat{\mathbf{E}}_{ij}^\dagger \cdot \hat{\mathbf{E}}_{ij} + \hat{\mathbf{D}}_{ij}^\dagger \cdot \hat{\mathbf{D}}_{ij} - 3 \right) . \label{spinproducttripletfermion}
\end{align}
\end{subequations}
As argued in Ref.~\onlinecite{PhysRevB.80.064410}, due to the sign of ferromagnetic (antiferromagnetic) interaction, decomposition of spin product into singlet (triplet) terms is unstable within an auxiliary field decoupling scheme. Therefore, for ferromagnetic interaction $J_{ij}<0$ (antiferromagnetic interaction $J_{ij}>0$), we write the spin product as \eqref{spinproducttripletfermion} (\eqref{spinproductsingletfermion}), and then perform a mean field decoupling to obtain a Hamiltonian quadratic in $f$-operators.\cite{PhysRevB.85.224428} With the constraint \eqref{numberconstraintfermion1} and \eqref{numberconstraintfermion2} enforced by introducing chemical potentials $a^l_i \in \mathbb{R}, l=1,2,3$, and some constant terms dropped, the mean field Hamiltonian reads
\begin{equation} \label{heisenbergmeanfieldfermion}
\begin{aligned}[b]
H_{\mathrm{MF}} =& - \frac{1}{4} \sum_{J_{ij}>0} J_{ij} \left[ \left( \chi_{ij}^* \hat{\chi}_{ij} + \mathrm{h.c.} - \left\lvert \chi_{ij} \right\rvert^2 \right) \right. \\
& \qquad \qquad \quad \: \: + \left. \left( {\Delta}_{ij}^* \hat{\Delta}_{ij} + \mathrm{h.c.} - \left\lvert \Delta_{ij} \right\rvert^2 \right) \right] \\
& + \frac{1}{4} \sum_{J_{ij}<0} J_{ij} \left[ \left( \mathbf{E}_{ij}^* \cdot \hat{\mathbf{E}}_{ij} + \mathrm{h.c.} - \left\lvert \mathbf{E}_{ij} \right\rvert^2 \right) \right. \\
& \qquad \qquad \quad \: \: + \left. \left( {\mathbf{D}}_{ij}^* \cdot \hat{\mathbf{D}}_{ij} + \mathrm{h.c.} - \left\lvert \mathbf{D}_{ij} \right\rvert^2 \right) \right] \\
& + \sum_i a^3_i \left( \sum_\alpha f_{i \alpha}^\dagger f_{i \alpha} - 1 \right) \\
& + \left( \sum_i (a^1_i+ia^2_i) f_{i \downarrow} f_{i \uparrow} + \mathrm{h.c.} \right) .
\end{aligned}
\end{equation}
Extremizing the mean field energy with respect to the variational parameters yields the following self consistent equations
\begin{equation} \label{selfconsistentequationfermion}
\frac{\partial \langle H_\mathrm{MF} \rangle}{\partial \mathcal{O}_{ij}} = 0 \iff \mathcal{O}_{ij} = \langle \hat{\mathcal{O}}_{ij} \rangle , \; \mathcal{O}_{ij} = \chi_{ij}, \Delta_{ij}, E^a_{ij}, D^a_{ij} ,
\end{equation}
while the chemical potentials $a^l_i$ are solved such that the constraints \eqref{numberconstraintaveragefermion1} and \eqref{numberconstraintaveragefermion2} are satisfied. In practice, the self consistent equations \eqref{selfconsistentequationfermion} are solved iteratively in momentum space.

\section{\label{fermionicpsganalysis} Projective Symmetry Group Analysis}
We are interested in the symmetric spin liquid states in non-symmorphic Kagome lattice. Using the method of projective symmetry group (PSG),\cite{PhysRevB.65.165113,PhysRevB.83.224413,1610.06191} we can identify all possible spin liquid ansatzes that respect the relevant symmetries. The main idea is that the mean field ansatzes for distinct spin liquid states are invariant under symmetry transformations followed by different gauge transformations. Therefore, spin liquid states can be distinguished by these different gauge transformations. \cite{PhysRevB.65.165113}

In order to treat the spin-singlet and triplet terms on equal footing, we closely follow Ref.~\onlinecite{1610.06191} and introduce
\begin{equation}
\Psi_i = \begin{pmatrix} f_{i \uparrow} & f_{i \downarrow}^\dagger \\ f_{i \downarrow} & -f_{i \uparrow}^\dagger \end{pmatrix}.
\end{equation}
We can then express the mean field Hamiltonian \eqref{heisenbergmeanfieldfermion} in the following form \cite{1610.06191}
\begin{equation} \label{heisenbergmeanfieldfermiontrace}
H_\mathrm{MF} = H^0 + H^x + H^y + H^z + C ,
\end{equation}
where
\begin{equation} \label{heisenbergmeanfieldfermiontracecomponent}
H^a = \pm \frac{1}{4} \sum_{ij} J_{ij} \left[ \mathrm{Tr} \left(\sigma^a \Psi_i U_{ij}^a \Psi_j^\dagger \right) + \frac{1}{2} \mathrm{Tr} \left( U_{ij}^{a \dagger} U_{ij}^a \right) \right]
\end{equation}
with $+$ sign for $a=0$ and $-$ sign for $a=x,y,z$. $\sigma^0$ is the $2 \times 2$ identity matrix and $\sigma^{x,y,z}$ are the Pauli matrices, while the ansatzes are given by
\begin{subequations}
\begin{align}
U_{ij}^0 &= \begin{pmatrix} \chi_{ij} & - \Delta_{ij}^* \\ - \Delta_{ij} & - \chi_{ij}^* \end{pmatrix}, \, J_{ij}>0 , \\
U_{ij}^a &= \begin{pmatrix} E_{ij}^a & D_{ij}^{a*} \\ - D_{ij}^a & E_{ij}^{a*}\end{pmatrix}, \, J_{ij}<0, \, a=x,y,z .
\end{align}
\end{subequations}
The second term in the square brackets in \eqref{heisenbergmeanfieldfermiontracecomponent} is irrelevant for the PSG analysis so we can simply ignore it for the moment. Moreover, we introduce $u_{ij}^a \equiv \pm J_{ij} U_{ij}^a /4$ to simplify the notation.

We also write the constraint $C$ explicitly in the Hamiltonian \eqref{heisenbergmeanfieldfermiontrace}, which is often referred to as on-site term. It has the form
\begin{equation} \label{onsitefermion}
\begin{aligned}[b]
C &= \sum_i \mathrm{Tr}\left(\Psi_i \begin{pmatrix} a^3_i & a^1_i - i a^2_i \\ a^1_i + i a^2_i & -a^3_i \end{pmatrix} \Psi_i^\dagger\right) \\
&= \sum_i \mathrm{Tr}\left(\Psi_i \sum_{l=1}^3 a^l_i \tau^l \Psi_i^\dagger\right)
\end{aligned}
\end{equation}
up to some multiplicative constant, which can be absorbed into the definition of chemical potentials $a^l_i$.

An $SU(2)$ spin rotation $R=\exp(i\bm{\sigma}\cdot\hat{\mathbf{n}}\theta/2)$ acts on $\Psi_i$ from the left by $\Psi_i \longrightarrow R^\dagger \Psi_i$.\cite{1610.06191} We see from \eqref{heisenbergmeanfieldfermiontracecomponent} that, while $H^0$ preserves global spin rotation symmetry of the original Heisenberg Hamiltonian \eqref{heisenberggeneral}, $H^{x,y,z}$ generically breaks it. We also observe that the mean field Hamiltonian \eqref{heisenbergmeanfieldfermiontrace} has the following $SU(2)$ gauge redundancy
\begin{subequations}
\begin{align}
G: \, & \Psi_i \longrightarrow \Psi_i G(i) , \label{gaugetransformationfermion} \\
& u_{ij}^a \longrightarrow G(i)^\dagger u_{ij}^a G(j) . \label{gaugetransformationansatzfermion}
\end{align}
\end{subequations}
In \eqref{gaugetransformationfermion}, we say that a gauge transformation $G(i) \in SU(2)$ acts on $\Psi_i$ from the right. The matrix $\tilde{\Psi}_i \equiv \Psi_i G(i)$ contains the transformed fermionic operators $\tilde{f}_{i \alpha}$, which leave the representation \eqref{spinoperator} of spin invariant, satisfying the same anticommutation relations \eqref{anticommutation} and constraints \eqref{numberconstraintfermion1} and \eqref{numberconstraintfermion2} as $f_{i \alpha}$, so it describes the same fermionic object. Therefore, ansatzes which differ by a gauge transformation give rise to the same physical state.

The presence of triplet terms in the mean field Hamiltonian breaks the global spin rotation symmetry. The remaining symmetries are the space group of non-symmorphic Kagome lattice and time reversal symmetry. Now, we are going to use the PSG to identify all possible spin liquid ansatzes that respect these symmetries.

Let $X$ be a space group element. In general, there will be a site independent spin rotation $R_X \in SU(2)$ associated with $X$, such that $X$ acts on $\Psi_i$ by \cite{1610.06191}
\begin{equation} \label{spinrotationspacegroup}
X: \Psi_i \longrightarrow R^\dagger_X \Psi_{X(i)} ,
\end{equation}
which changes the mean field Hamiltonian by
\begin{equation} \label{heisenbergmeanfieldfermiontracecomponentchange}
H^a \overset{X}{\longrightarrow} \sum_{ij} \mathrm{Tr} \left(R_X \sigma^a R^\dagger_X \Psi_{X(i)} u_{ij}^a \Psi_{X(j)}^\dagger \right) .
\end{equation}
Let us first inspect the simplest case $a=0$, in which $R_X \sigma^0 R^\dagger_X=1$. We want the Hamiltonian to be invariant under the action of $X$. By the $SU(2)$ gauge redundancy, this requires $u^0_{ij}$ to be equal to $u^0_{X(i)X(j)}$ up to a gauge transformation $G_X(\mathbf{r}) \in SU(2)$,
\begin{equation} \label{singletansatzrelationfermion}
u^0_{X(i)X(j)} = G_X(X(i)) u^0_{ij} G_X^\dagger(X(j)) .
\end{equation}
This suggests that we can view the action of $X$ and $G_X$ on $u^0_{ij}$ as \cite{1610.06191}
\begin{subequations}
\begin{align}
X &: u^0_{ij} \longrightarrow u^0_{X^{-1}(i)X^{-1}(j)} , \label{singletansatspacegroupactionfermion} \\
G_X &: u^0_{ij} \longrightarrow G_X(X(i)) u^0_{ij} G_X^\dagger(X(j)) , \label{singletansatzgaugeactionfermion}
\end{align}
\end{subequations}
while the compound operator $G_X X$ acts trivially on $u^0_{ij}$. The set of all $G_X X$ that leaves the ansatz invariant is defined as projective symmetry group (PSG). This definition includes $X=\mathcal{T}$, the time reversal operator, whose action will be discussed later.

When $a=x,y,z$ in \eqref{heisenbergmeanfieldfermiontracecomponentchange}, we have to take into account the effect of spin rotation associated with the space group element. We can map the $SU(2)$ rotation to an $SO(3)$ rotation on the Pauli matrices,\cite{1610.06191}
\begin{equation} \label{su2so3map}
R_X \sigma^a R^\dagger_X = \sum_{b=x,y,z} O_X^{ab} \sigma^b ,
\end{equation}
such that the triplet Hamiltonian changes by
\begin{equation}
\sum_{a=x,y,z} H^a \overset{X}{\longrightarrow} \sum_{ij} \sum_{ab} O_X^{ab} \mathrm{Tr} \left(\sigma^b \Psi_{X(i)} u_{ij}^a \Psi_{X(j)}^\dagger \right) .
\end{equation}
By $SU(2)$ gauge redundancy, and the fact that $O^{ab} = (O^\mathrm{T})^{ba} = (O^{-1})^{ba}$ for any $O \in SO(3)$, we must have
\begin{equation*}
u_{X(i)X(j)}^b = \sum_{a} (O^{-1}_X)^{ba} G_X(X(i)) u_{ij}^a G_X^\dagger(X(j)) ,
\end{equation*}
or
\begin{equation} \label{tripletansatzrelationfermion}
\sum_{b} O_X^{ab} u_{X(i)X(j)}^b = G_X(X(i)) u_{ij}^a G_X^\dagger(X(j)), \, a=x,y,z.
\end{equation}
Notice that reflection is not a proper rotation and cannot be described by an $SO(3)$ matrix. We can however generalize $SO(3)$ to $O(3)$ to include improper rotations. In particular, we will see that the action of glide, which is the combination of reflection and half translation, in spin space is described by an $O(3)$ matrix with $\det = -1$.

Consider now the antiunitary time reversal operator,
\begin{equation}
\mathcal{T} = i \sigma^y K
\end{equation}
where $i \sigma^y$ operates on $\Psi_i$ from the left and $K$ complex conjugates any number on the right. It acts on the Hamiltonian as \cite{1610.06191}
\begin{equation} \label{timereversalactionhamiltonianfermion}
\begin{aligned}[b]
\mathcal{T} H^a \mathcal{T}^{-1} &= \sum_{ij} \mathrm{Tr} \left(\sigma^{a*} i\sigma^y \Psi_{X(i)} K u_{ij}^a \Psi_{X(j)}^\dagger K (-i\sigma^y) \right) \\
&= \mathrm{Tr} \left(\sigma^y \sigma^{a*} \sigma^y \Psi_{X(i)} u_{ij}^{a*} \Psi_{X(j)}^\dagger \right) \\
&\longrightarrow \mathrm{Tr} \left(\sigma^y \sigma^{a*} \sigma^y \Psi_{X(i)} (i \sigma^y) u_{ij}^{a*} (-i\sigma^y) \Psi_{X(j)}^\dagger \right) \\
&= \mathrm{Tr} \left(\sigma^{a} \Psi_{X(i)} (-u_{ij}^{a}) \Psi_{X(j)}^\dagger \right) .
\end{aligned}
\end{equation}
where in the third line we have introduced a gauge transformation $i \tau^2$,\cite{PhysRevB.65.165113} and in the last line we have used $\sigma^y \sigma^{0*} \sigma^y = \sigma^0$, $\sigma^y u_{ij}^{0*} \sigma^y = -u_{ij}^0$, $\sigma^y \sigma^{a*} \sigma^y = -\sigma^a$, $\sigma^y u_{ij}^{a*} \sigma^y = u_{ij}^a$ for $a=x,y,z$. \cite{1610.06191} The Hamiltonian must be invariant under the action of $\mathcal{T}$ if the system has time reversal symmetry, which requires
\begin{equation} \label{timereversalansatzfermion}
- u_{ij}^a = G_\mathcal{T}(i) u_{ij}^a G_\mathcal{T}^\dagger(j),\, a=0,x,y,z,
\end{equation}
for some $G_\mathcal{T}(\mathbf{r}) \in SU(2)$. Note that both $\sigma^a$ and $\tau^a$ are identity ($a=0$) or Pauli matrices ($a=x,y,z$ or $1,2,3$), but $\sigma^a$ denotes spin rotation while $\tau^a$ denotes gauge transformation.

Suppose that $G_X X \in \mathrm{PSG}$. Applying a gauge transformation $G(i) \in SU(2)$ on the ansatz, $G_X$ changes as $G_X(i) \longrightarrow G(i) G_X(i) G^\dagger(X^{-1}(i))$ such that $G_X X$ is still a PSG element.\cite{PhysRevB.83.224413} A subgroup of PSG known as invariant gauge group (IGG) consists of elements of the form $G_I I$ where $I$ is the identity element. It is the group of pure gauge transformation that leave the ansatz invariant.\cite{PhysRevB.65.165113,PhysRevB.83.224413} Typically, when both hopping and pairing terms are present in the mean field Hamiltonian, the IGG is just $\mathbb{Z}_2 = \lbrace - 1, + 1 \rbrace$.

\subsection{\label{algebraicpsgfermion} Algebraic PSG}
The algebraic relations among the symmetry operations (space group and time reversal) constrain the possible forms of the gauge transformations $G_X$. In particular, the algebraic identities \eqref{algebraicrelation1} $-$ \eqref{algebraicrelation10} impose a set of consistent conditions \eqref{algebraicconstraint1} $-$ \eqref{algebraicconstraint13} on $G_X$. These consistent conditions (or algebraic constraints) are the same for the singlet and triplet ansatzes, since there is no net spin rotation for symmetry operations that amount to identity. We do have to be careful not to neglect the effect of spin rotation when relating the triplet ansatzes by symmetry, which we will soon discuss. The PSG in which gauge transformation $G_X$ associated with symmetry $X$ satisfies the algebraic constraints is known as algebraic PSG.

The final result of algebraic PSG is
\begin{subequations}
\begin{align}
G_{T_x}(x,y,s) &= \tau^0 , \label{translationxpsgfermion} \\
G_{T_y}(x,y,s) &= \tau^0 , \label{translationypsgfermion} \\
G_{C_2}(x,y,s) &= \eta_{C_2T_y}^{x+y} g_{C_2}(s) , \\
G_h(x,y,s) &= \eta_{C_2T_y}^y g_h(s) , \\
G_\mathcal{T}(x,y,s) &= i \tau^2 , \label{timereversalpsgfermion}
\end{align}
\end{subequations}
with $g_X(s) \equiv G_X(0,0,s)$ for $X=C_2,h$ given in TABLE \ref{fermionicpsgsolution}. Consistent combinations of the $\mathbb{Z}_2$ variables $\eta_{C_2}$, $\eta_{C_2T_y}$, $\eta_{C_2\mathcal{T}}$ and $\eta_{h\mathcal{T}}$, which arise from the algebraic constraints \eqref{algebraicconstraint1} $-$ \eqref{algebraicconstraint13}, result in various forms of $g_{C_2}(s)$ and $g_h(s)$, which in turn give rise to twelve distinct $\mathbb{Z}_2$ spin liquid states that respect the space group of non-symmorphic Kagome lattice and time reversal symmetry. Detailed derivation of the algebraic PSG can be found in Appendix \ref{fermionicpsgderive}.

\begin{table*}
\caption{\label{fermionicpsgsolution} The possible forms of $g_X(s) \equiv G_X(0,0,s)$ for $X=C_2,h$, which lead to twelve distinct $\mathbb{Z}_2$ spin liquid states. These states are labeled by a number $1,\ldots,6$ followed by $a$ when $\eta_{C_2T_y}=+1$, or $b$ when $\eta_{C_2T_y}=-1$.}
\begin{ruledtabular}
\begin{tabular}{c|cccccccccccc}
$ \mathrm{No.} $ & $g_{C_2}(1)$ & $g_{C_2}(2)$ & $g_{C_2}(3)$ & $g_{C_2}(4)$ & $g_{C_2}(5)$ & $g_{C_2}(6)$ & $g_{h}(1)$ & $g_{h}(2)$ & $g_{h}(3)$ & $g_{h}(4)$ & $g_{h}(5)$ & $g_{h}(6)$ \\ \hline
$ 1 $ & $\eta_{C_2T_y} \tau^0$ & $\eta_{C_2T_y} \tau^0$ & $\tau^0$ & $\tau^0$ & $\tau^0$ & $\tau^0$ & $\eta_{C_2T_y}\tau^0$ & $\tau^0$ & $\tau^0$ & $\tau^0$ & $\tau^0$ & $\tau^0$ \\
$ 2 $ & $\eta_{C_2T_y} \tau^0$ & $\eta_{C_2T_y} \tau^0$ & $\tau^0$ & $\tau^0$ & $\tau^0$ & $\tau^0$ & $\eta_{C_2T_y} (-i \tau^3)$ & $i \tau^3$ & $-i \tau^3$ & $i \tau^3$ & $-i \tau^3$ & $i \tau^3$ \\
$ 3 $ & $\eta_{C_2T_y} (i \tau^2)$ & $\eta_{C_2T_y} (i \tau^2)$ & $i \tau^2$ & $i \tau^2$ & $i \tau^2$ & $-i \tau^2$ & $-\eta_{C_2T_y} \tau^0$ & $-\tau^0$ & $\tau^0$ & $\tau^0$ & $\tau^0$ & $\tau^0$ \\
$ 4 $ & $\eta_{C_2T_y} (i \tau^2)$ & $\eta_{C_2T_y} (i \tau^2)$ & $i \tau^2$ & $i \tau^2$ & $i \tau^2$ & $i \tau^2$ & $\eta_{C_2T_y} (i \tau^3)$ & $-i \tau^3$ & $-i \tau^3$ & $-i \tau^3$ & $i \tau^3$ & $i \tau^3$ \\
$ 5 $ & $\eta_{C_2T_y}(i \tau^3)$ & $\eta_{C_2T_y}(i \tau^3)$ & $i \tau^3$ & $i \tau^3$ & $i \tau^3$ & $-i \tau^3$ & $-\eta_{C_2T_y} \tau^0$ & $-\tau^0$ & $\tau^0$ & $\tau^0$ & $\tau^0$ & $\tau^0$ \\
$ 6 $ & $\eta_{C_2T_y}(i \tau^3)$ & $\eta_{C_2T_y}(i \tau^3)$ & $i \tau^3$ & $i \tau^3$ & $i \tau^3$ & $-i \tau^3$ & $\eta_{C_2T_y}(i \tau^3)$ & $-i \tau^3$ & $i \tau^3$ & $i \tau^3$ & $-i \tau^3$ & $-i \tau^3$
\end{tabular}
\end{ruledtabular}
\end{table*}

\subsection{\label{meanfieldansatzfermion} Mean Field Ansatz}

There are four different exchange interactions in the spin model of Volborthite, as discussed in Section \ref{latticehamiltonian}. For $\mathbb{Z}_2$ spin liquid states, both hopping and pairing channel are present for every exchange interaction. Each singlet channel contributes one mean field parameter, while each triplet channel contributes three mean field parameters, which correspond to its $x,y,z$ components. Therefore, we have $16$ independent mean field parameters, which we denote by $\chi$, $\Delta$, $\mathbf{E}'$, $\mathbf{D}'$, $\mathbf{E}_1$, $\mathbf{D}_1$, $\chi_2$ and $\Delta_2$, depending on which exchange couplings they are associated with. We also consider (for reasons that will be explained in Section \ref{result}) the root $U(1)$ states, which are described by the Hamiltonian \eqref{heisenbergmeanfieldfermion} without all the pairing terms. Therefore, the root $U(1)$ states have $8$ independent mean field parameters.

All these mean field parameters are complex-valued in general. However, through \eqref{timereversalansatzfermion} and \eqref{timereversalpsgfermion}, time reversal symmetry restricts the singlet parameters $\chi_{ij}$ and $\Delta_{ij}$ to be real, and the triplet parameters $\mathbf{E}_{ij}$ and $\mathbf{D}_{ij}$ to be imaginary, for $u_{ij}^a$ to be nonzero.

Symmetry-related singlet ansatzes $u^0_{ij}$ are generated by \eqref{singletansatzrelationfermion}, while triplet ansatzes $u^a_{ij}$ by \eqref{tripletansatzrelationfermion}. The matrix $O_X \in SO(3)$ in \eqref{su2so3map} that describes the spin rotation associated with space group element $X$ \eqref{spinrotationspacegroup} is trivial for translations $X=T_x,T_y$. For $\pi$-rotation and glide, we have
\begin{subequations}
\begingroup
\allowdisplaybreaks
\begin{align}
O_{C_2} &= \begin{pmatrix} -1 & 0 & 0 \\ 0 & -1 & 0 \\ 0 & 0 & 1 \end{pmatrix} , \\
O_h &= \begin{pmatrix} 1 & 0 & 0 \\ 0 & -1 & 0 \\ 0 & 0 & 1 \end{pmatrix} .
\end{align}
\endgroup
\end{subequations}
While $O_{C_2} \in SO(3)$, we see that $O_h$ having $\det = -1$ is an element in $O(3) \setminus SO(3)$.

As in the bosonic PSG,\cite{1702.04360} the gauge matrix \eqref{translationxpsgfermion} and \eqref{translationypsgfermion} of lattice translations in fermionic PSG are trivial. To construct $H_\mathrm{MF}$, it is therefore sufficient to relate the mean field parameters by \eqref{singletansatzrelationfermion} or \eqref{tripletansatzrelationfermion} in the $(0,0)$ unit cell, because the same set of relations hold in all other unit cells. In other words, the mean field ansatz does not go beyond the physical unit cell. We explained in Ref.~\onlinecite{1702.04360} that this is a consequence of non-symmorphic symmetry. The proof, which involves only few steps of simple algebraic manipulation, will however be repeated here for completeness.

Consider the algebraic constraints \eqref{algebraicconstraint5} and \eqref{algebraicconstraint7}, which originate from the spatial identities \eqref{algebraicrelation5} and \eqref{algebraicrelation7} respectively. Abbreviating $G_X X$ as $\tilde{X}$, we have
\begin{equation} \label{translationfractionalization}
\begin{aligned} [b]
{\tilde{T}_x}^{-1} {\tilde{T}_y}^{-1} \tilde{T}_x \tilde{T}_y &= (\eta_{h} \tilde{h}^2)^{-1} {\tilde{T}_y}^{-1} \eta_{h} \tilde{h}^2 \tilde{T}_y \\
&= \tilde{h}^{-1} \left(\tilde{h}^{-1} {\tilde{T}_y}^{-1} \tilde{h} {\tilde{T}_y}^{-1}\right) \tilde{T}_y \tilde{h} \tilde{T}_y \\
&= \eta_{hT_y} \eta_{hT_y} \\
&= +1 .
\end{aligned}
\end{equation}
\eqref{translationfractionalization} rules that the mean field ansatz allowed by PSG can never go beyond the physical unit cell, as it would require $ {\tilde{T}_x}^{-1} {\tilde{T}_y}^{-1} \tilde{T}_x \tilde{T}_y = -1 $ should the contrary be true. The same argument holds for three other non-symmorphic plane crystallographic groups $pg$, $p2mg$ and $p4gm$. In contrast, the algebraic PSG of isotropic Kagome lattice, where non-symmorphic symmetry is absent, allows certain spin liquid states to have mean field ansatz that enlarges the physical unit cell.\cite{PhysRevB.83.224413}

PSG analysis also requires that the on-site chemical potential $\sum_l a^l_i \tau^l \equiv u_{ii}^0$ in \eqref{onsitefermion} satisfies
\begin{subequations}
\begin{align}
u^0_{X(i)X(i)} &= G_X(X(i)) u^0_{ii} G_X^\dagger(X(i)), \, X \in \lbrace T_x,T_y,C_2,h \rbrace , \label{onsiterelationspacegroupfermion} \\
- u^0_{ii} &= G_\mathcal{T}(i) u^0_{ii} G_\mathcal{T}^\dagger(i) , \label{onsiterelationtimereversalfermion}
\end{align}
\end{subequations}
for consistency.\cite{PhysRevB.83.224413} By translational invariance, the chemical potential can only have sublattice dependence, so we write
$u_{ii}^0 \equiv \Lambda_s, s=1,\ldots,6$. Through \eqref{onsiterelationtimereversalfermion} and \eqref{timereversalpsgfermion}, time reversal symmetry restricts the chemical potential to be
\begin{equation}
\Lambda_s = \mu_s \tau^3 + \nu_s \tau^1 ,
\end{equation}
for some $\mu_s, \nu_s \in \mathbb{R}$. The remaining symmetries, $C_2$ and $h$, further imply that there can only be at most two independent $\mu_s$ and $\nu_s$, so in general we have four chemical potentials.

We will be interested in the fermionic states 3a, 3b, 4a and 4b, which can be connected to Schwinger boson spin liquid states, as discussed in Section \ref{bosonfermionmap}. In these states, we find only one independent $\mu_s$ and $\nu_s$, so the number of chemical potentials is reduced to two, which we argue as follows. Since $G_\mathcal{T}(x,y,s)= i\tau^2$, by \eqref{onsiterelationtimereversalfermion} we have
\begin{equation*}
-\Lambda_s = \tau^2 \Lambda_s \tau^2
\end{equation*}
for all $s$. Furthermore, $g_{C_2}(s) = \pm i \tau^2$ (see TABLE \ref{fermionicpsgsolution}) for 3a, 3b, 4a and 4b, by \eqref{onsiterelationspacegroupfermion} we have
\begin{equation*}
\begin{aligned}
\Lambda_1 &= \tau^2 \Lambda_5 \tau^2 = -\Lambda_5 , \\
\Lambda_2 &= \tau^2 \Lambda_4 \tau^2 = -\Lambda_4 , \\
\Lambda_3 &= \tau^2 \Lambda_3 \tau^2 = -\Lambda_3 , \\
\Lambda_6 &= \tau^2 \Lambda_6 \tau^2 = -\Lambda_6 ,
\end{aligned}
\end{equation*}
which implies $\Lambda_3=\Lambda_6=0$. We can further use $h$ to relate the chemical potentials at $s=1,2,4,5$, which for brevity we show the final result directly,
\begin{equation*}
\begin{aligned}
& \mu_1 = \mu_2 = - \mu_4 = - \mu_5, \, \nu_1 = \nu_2 = - \nu_4 = - \nu_5, \, \textrm{for 3a and 3b}; \\
& \mu_1 = \mu_2 = - \mu_4 = - \mu_5, \, \nu_1 = -\nu_2 = \nu_4 = - \nu_5, \, \textrm{for 4a and 4b}.
\end{aligned}
\end{equation*}
Therefore, there is only one independent $\mu_s$ and $\nu_s$ for the fermionic states 3a, 3b, 4a and 4b, which simplifies the calculation.

\section{\label{bosonfermionmap} Mapping between Bosonic and Fermionic Spin Liquid States}

Lu \textit{et al.} developed a framework to connect bosonic and fermionic spin liquid states on the isotropic Kagome lattice through vison PSG in Ref.~\onlinecite{1403.0575}. Similar analysis has been carried out in square lattice,\cite{PhysRevB.94.035160} rectangular lattice,\cite{1603.03041} and breathing Kagome lattice.\cite{1605.05322} We would like to do the same for non-symmorphic Kagome lattice, to figure out the correspondence between the eight bosonic states and the twelve fermionic states resulting from PSG analysis in Ref.~\onlinecite{1702.04360} and Section \ref{fermionicpsganalysis}, respectively. The bosonic states $(p_2,p_3,p_{13})$ are labeled by three $\mathbb{Z}_2$ variables. In particular, we would like to know the fermionic states that are connected to the most energetically favorable bosonic state $(1,0,0)$ and the highly competing $(1,1,0)$, from which a $(q,0)$ spiral order and a $(\pi,\pi)$ spin density wave develop, respectively, upon spinon condensation.

The mapping is based on the idea of symmetry fractionalization and fusion rule, which we briefly explain below following Ref.~\onlinecite{1403.0575}. Spinons and visons are collectively known as anyons. Any local excitation in a $\mathbb{Z}_2$ spin liquid, which must be a bound state of two anyons of the same type, can only gain a trivial phase factor ($+1$) under symmetry operations that amount to identity. Therefore, the phase factor picked up by one anyon under these symmetry operations is quantized to be $\pm 1$. This is known as symmetry fractionalization, and the phase factor $\pm 1$ is known as symmetry quantum number. The three types of anyon, bosonic spinon $b$, fermionic spinon $f$ and vison $v$, obey the following abelian fusion rule,
\begin{equation*}
\begin{split}
& b \times f = v, \, f \times v = b, \, v \times b = f, \\
& b \times b = f \times f = v \times v = 1,
\end{split}
\end{equation*}
where $1$ represents local excitations carrying integer spin. The fusion rule tells us that vison is a bound state of bosonic spinon and fermionic spinon, etc. Let $X$ be a string of symmetry operations that amount to identity. Suppose that under the action of $X$, the phase gain of bosonic spinon, fermionic spinon and vison are $\phi_b$, $\phi_f$ and $\phi_v$ respectively. Then, by the fusion rule, they satisfy
\begin{equation}
e^{i\phi_f} = e^{i\phi_t} e^{i\phi_b} e^{i\phi_v} ,
\end{equation}
where we have introduced the twist factor $e^{\phi_t}=\pm 1$. When $e^{\phi_t}=+1$ ($e^{\phi_t}=-1$), the fusion rule is said to be trivial (nontrivial). Nontrivial fusion rule arises due to mutual semion satistics. In short, $e^{i\phi_t}=-1$ when anyon of one type, say $b$, encircles anyon of another type, say $v$, under the action of $X$.

\begin{table}
\caption{\label{bosonfermionmaptable} Algebraic identities and correspondence between bosonic spinon, fermionic spinon and vison PSG on non-symmorphic Kagome lattice.}
\begin{ruledtabular}
\begin{tabular}{c|c|c|c|c}
algebraic identity & boson & fermion & vison & trivial fusion rule? \\ \hline
$T_x^{-1} T_y^{-1} T_x T_y$ & $1$ & $1$ & $1$ & Yes \\
$C_2^2$ & $\left(-1\right)^{p_2}$ & $\eta_{C_2}$ & $1$ & No \\
$C_2 T_x C_2^{-1} T_x$ & $\left(-1\right)^{p_3}$ & $\eta_{C_2T_y}$ & $-1$ & Yes \\
$C_2 T_y C_2^{-1} T_y$ & $\left(-1\right)^{p_3}$ & $\eta_{C_2T_y}$ & $-1$ & Yes \\
$T_x^{-1} h^2$ & $1$ & $1$ & $-1$ & No \\
$h^{-1} T_x^{-1} h T_x$ & $1$ & $1$ & $1$ & Yes \\
$h^{-1} T_y h T_y$ & $\left(-1\right)^{p_3}$ & $\eta_{C_2T_y}$ & $-1$ & Yes \\
$T_x T_y h^{-1} C_2 h C_2$ & $1$ & $1$ & $-1$ & No \\
$\mathcal{T}^2$ & $-1$ & $-1$ & $1$ & Yes \\
$T_x^{-1}\mathcal{T}^{-1}T_x\mathcal{T}$ & $1$ & $1$ & $1$ & Yes \\
$T_y^{-1}\mathcal{T}^{-1}T_y\mathcal{T}$ & $1$ & $1$ & $1$ & Yes \\
$C_2^{-1}\mathcal{T}^{-1}C_2\mathcal{T}$ & $\left(-1\right)^{p_2+p_3}$ & $\eta_{C_2\mathcal{T}}$ & $1$ & No \\
$h^{-1}\mathcal{T}^{-1}h\mathcal{T}$ & $\left(-1\right)^{p_{13}}$ & $\eta_{h\mathcal{T}}$ & $1$ & No \\
\end{tabular}
\end{ruledtabular}
\end{table}

The symmetry quantum numbers of bosonic and fermionic spinons are known from solving their respective algebraic PSG (see Ref.~\onlinecite{1702.04360} and Appendix \ref{fermionicpsgderive}). Therefore, the remaining ingredients required to complete the mapping between bosonic and fermionic spin liquid states are the vison PSG and the various twist factors, which we provide in Appendix \ref{visonpsgderive} and \ref{fusionrule}. The correspondence between bosonic spinon, fermionic spinon and vison PSG on the non-symmorphic Kagome lattice is shown in TABLE \ref{bosonfermionmaptable}, from which we can easily identify
\begin{subequations}
\begin{align}
(-1)^{p_2} &= - \eta_{C_2} , \\
(-1)^{p_3} &= - \eta_{C_2 T_y} , \\
(-1)^{p_2+p_3} &= - \eta_{C_2 \mathcal{T}} , \\
(-1)^{p_{13}} &= - \eta_{h \mathcal{T}} .
\end{align}
\end{subequations}
This maps the bosonic state $(p_2,p_3,p_{13}) = (1,0,0)$, which gives rise to $(q,0)$ spiral order, to the fermionic state 4b, $(0,1,0)$ to 4a, $(1,0,1)$ to 3b, and $(0,1,1)$ to 3a. All other bosonic (fermionic) states do not have fermionic (bosonic) counterpart. For a given lattice, the number of fermion spin liquid states is usually greater than bosonic spin liquid states under PSG classification, which is true in both the isotropic Kagome lattice \cite{PhysRevB.74.174423,PhysRevB.83.224413} and the non-symmorphic Kagome lattice. However, all the bosonic states in isotropic Kagome lattice have fermionic correspondence,\cite{1403.0575} unlike the non-symmorphic Kagome lattice, where half of the bosonic states have no fermionic counterpart.

\section{\label{result} Spectrum of $\mathbb{Z}_2$ and $U(1)$ Fermionic States with Bosonic Correspondence}

In Section \ref{bosonfermionmap}, we showed that only four out of twelve fermionic spin liquid states have bosonic correspondence. They are 3a, 3b, 4a and 4b, which are connected to the bosonic spin liquid states $(0,1,1)$, $(0,1,0)$, $(1,0,1)$ and $(1,0,0)$ respectively. We narrow down our investigation to these four fermionic states and study their generic mean field spectrum. We consider the $\mathbb{Z}_2$ states as well as the corresponding root $U(1)$ states, because thermal Hall effect, which is observed in Volborthite,\cite{1608.00444} would arise in $U(1)$ spin liquid with spinon Fermi surface. As argued in Ref.~\onlinecite{PhysRevLett.104.066403,RevModPhys.78.17}, when the mean field spin liquid state has $U(1)$ gauge symmetry, the spinons are coupled to an emergent $U(1)$ gauge field. When an external magnetic field is applied, a fictitious magnetic field is induced, which in turn leads to an effective Lorentz force for the spinons and the finite thermal Hall conductivity arises when the spinons form a Fermi surface.

\begin{figure*}
\begin{tabular}{>{\centering\arraybackslash} m{8.5 cm} >{\centering\arraybackslash} m{8.5 cm}}
\includegraphics[scale=0.25]{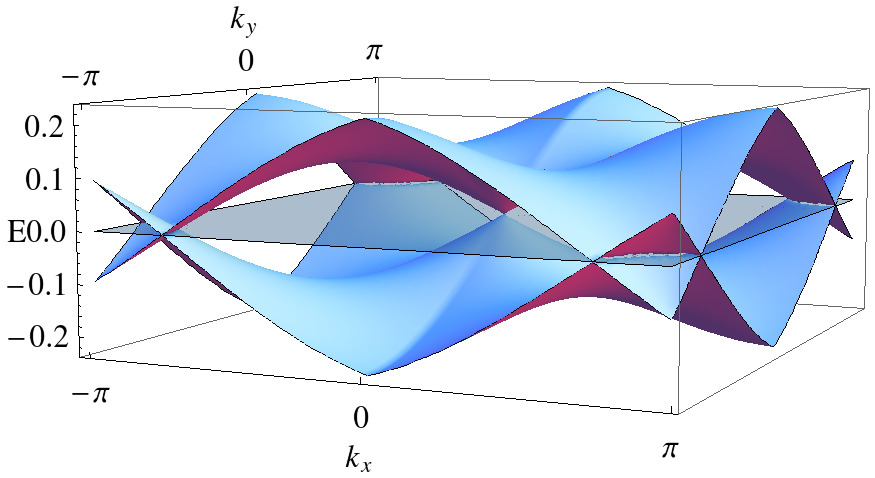} & \includegraphics[scale=0.25]{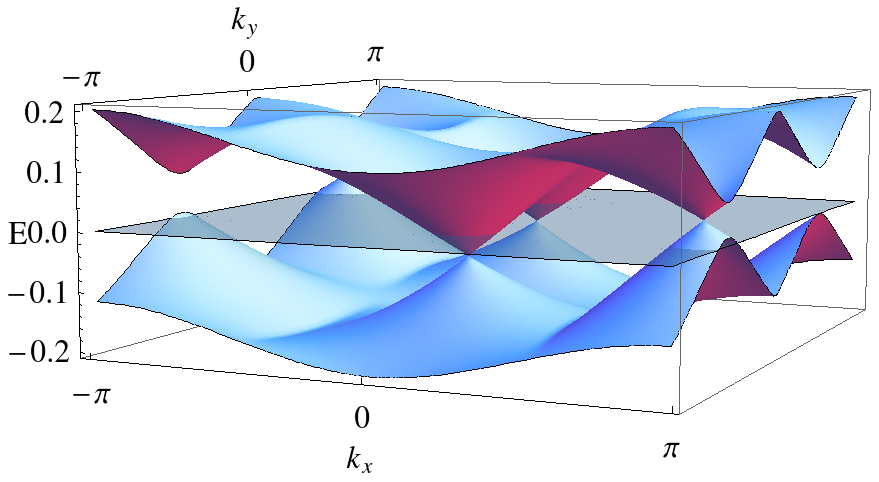} \\
(a) & (b) \\
\includegraphics[scale=0.25]{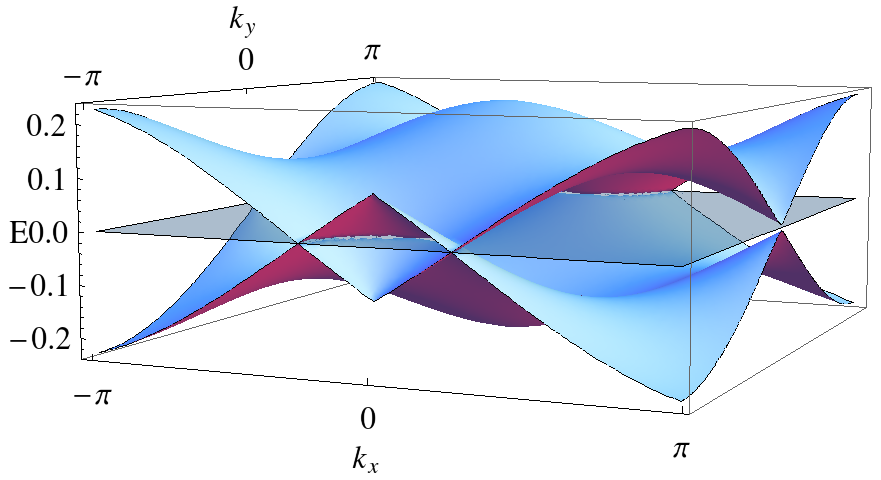} & \includegraphics[scale=0.25]{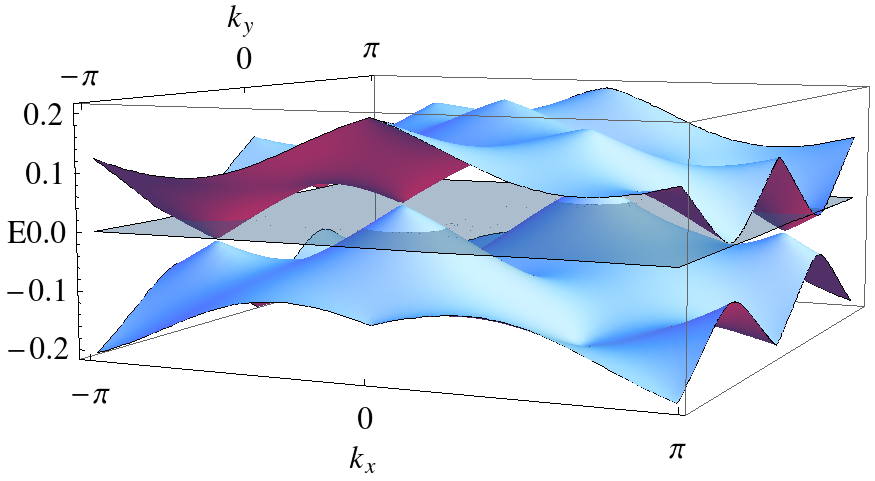} \\
(c) & (d)
\end{tabular}
\caption{\label{u1fermionicspectrum} Mean field spectrum around the Fermi level $E_\mathrm{F}=0$ (indicated by transparent sheet) of the root $U(1)$ state of (a) 3a, (b) 3b, (c) 4a and (d) 4b, plotted in the Brillouin zone $-\pi \leq k_x, k_y \leq \pi$. All these states have Fermi surface. Energy $E$ is in units of $J/4$ where $J$ is the strongest coupling constant.}
\end{figure*}

\begin{figure*}
\begin{tabular}{>{\centering\arraybackslash} m{8.5 cm} >{\centering\arraybackslash} m{8.5 cm}}
\includegraphics[scale=0.25]{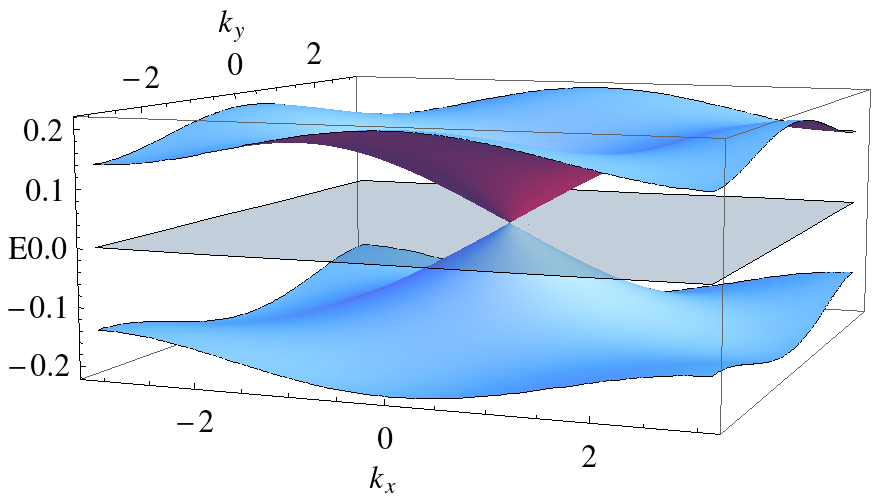} & \includegraphics[scale=0.25]{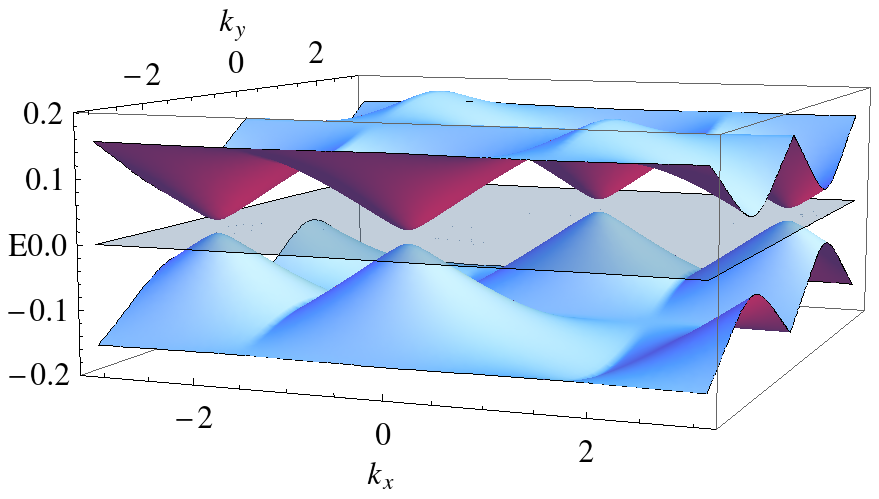} \\
(a) & (b) \\
\includegraphics[scale=0.25]{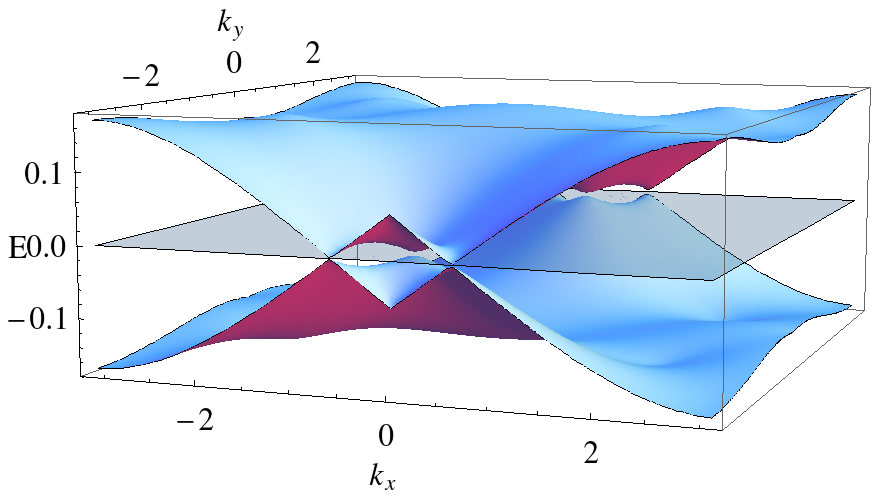} & \includegraphics[scale=0.25]{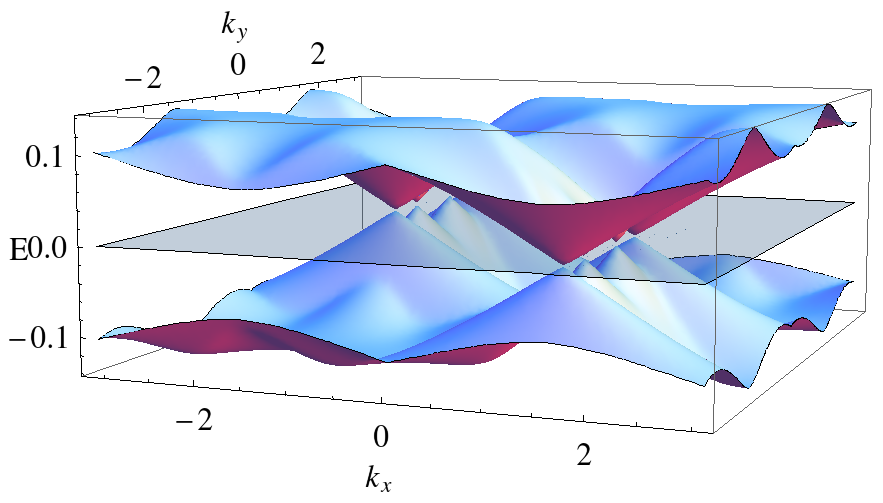} \\
(c) & (d)
\end{tabular}
\caption{\label{z2fermionicspectrum} Mean field spectrum around the Fermi level $E_\mathrm{F}=0$ (indicated by transparent sheet) of the $\mathbb{Z}_2$ spin liquid state (a) 3a, (b) 3b, (c) 4a and (d) 4b, plotted in the Brillouin zone $-\pi \leq k_x, k_y \leq \pi$. 3a, 4a and 4b have Dirac point(s), while the spectrum of 3b is gapped. Energy $E$ is in units of $J/4$ where $J$ is the strongest coupling constant.}
\end{figure*}

Our result is summarized in TABLE \ref{fermionicspectrumtable}. We find that the root $U(1)$ states of 3a, 3b, 4a and 4b all feature Fermi surfaces.
Hence these $U(1)$ spin liquid states may be possible candidates for the paramagnetic phase of Volborthite and provide an explanation of the finite thermal Hall conductivity seen above $1 \, \mathrm{K}$. We find that the spinon energy bands of these $U(1)$ states are doubly degenerate. For the $\mathbb{Z}_2$ spin liquids, we find that 3a, 4a and 4b have Dirac point(s) while 3b has a gapped spectrum. Here, the spinon energy bands of 3a and 4a are doubly degenerate, while those of 3b and 4b are non-degenerate. We plot the dispersion immediately above and below the Fermi level in the root $U(1)$ and $\mathbb{Z}_2$ states in FIG. \ref{u1fermionicspectrum} and \ref{z2fermionicspectrum}, respectively.

\begin{table}
\caption{\label{fermionicspectrumtable} $\mathbb{Z}_2$ and root $U(1)$ spectrum of the fermionic spin liquid states 3a, 3b, 4a and 4b from perturbed mean field solutions.}
\begin{tabular}{>{\centering\arraybackslash} m{1 cm} |>{\centering\arraybackslash} m{3.2 cm} |>{\centering\arraybackslash} m{2.5 cm}}
\hline \hline
state & root $U(1)$ spectrum & $\mathbb{Z}_2$ spectrum \\ \hline
3a & Fermi surface & Dirac point \\
3b & Fermi surface & Gapped \\
4a & Fermi surface & Dirac point \\
4b & Fermi surface & Dirac point \\
\hline \hline
\end{tabular}
\end{table}

\section{\label{summary} Discussion}

In this work, we focus on the paramagnetic state of Volborthite above the magnetic ordering temperature $T \approx 1 \, \mathrm{K}$, where a recent experiment \cite{1608.00444} reveals finite thermal Hall conductivity. Given that the magnetic ordering temperature is much smaller than the Curie-Weiss temperature of $140 \, \mathrm{K}$,\cite{ncomms1875} we take the point of view that the paramagnetic state right above $1 \, \mathrm{K}$ may be better considered as a finite temperature version of a proximate paramagnetic quantum state, which may be obtained by suppressing the magnetic order at zero temperature. Hence we are interested in possible quantum (paramagnetic) ground states that are closely related to the magnetic order below $1 \, \mathrm{K}$ and exhibit finite thermal Hall conductivity.

We notice that the microscopic spin model derived from DFT \cite{PhysRevLett.117.037206} suggests non-symmorphic symmetry of the underlying lattice structure, which is described by the plane group $p2gg$ and can be regarded as a distorted version of the Kagome lattice. As explained in the main text, this non-symmorphic Kagome lattice has six sites per unit cell with one electron per site in the insulating phase. According to a recent work \cite{PhysRevLett.117.096404} on the extension of the HOLSM theorem, a trivial band insulator is possible in this lattice only when the electron filling is $4n$, where $n \in \mathrm{N}$. Hence any state without broken symmetry in this system must be either topologically ordered (with gapped excitations) or gapless.\cite{nphys2600} Since Volborthite is an insulator, this means any (non-symmetry-breaking) paramagnetic quantum ground state must be nontrivial.

In this context, we investigate possible quantum spin liquid states in the non-symmorphic Kagome lattice. These states are nontrivial paramagnetic quantum ground states potentially relevant to Volborthite. In our previous study,\cite{1702.04360} we identified a $(q,0)$ spiral order and a $(\pi,\pi)$ spin density wave as the most promising magnetically ordered states for the magnetic order seen below $1 \, \mathrm{K}$ in Volborthite. These magnetically ordered states can be obtained from $\mathbb{Z}_2$ spin liquid phases with bosonic spinons, via condensation of the spinons. We showed that there exist eight possible bosonic spin liquid states and two of them are related to the $(q,0)$ spiral order and the $(\pi,\pi)$ spin density wave respectively.

While the bosonic spin liquid states mentioned above are closely related to the magnetic orders, these states are necessarily gapped (topologically ordered) and cannot naturally explain the finite thermal Hall conductivity. Hence, in the current work, we study quantum spin liquid states with fermionic spinons, which allow both gapped and gapless excitation spectra. We find that there are twelve distinct $\mathbb{Z}_2$ fermionic spin liquid states. Only four of the twelve $\mathbb{Z}_2$ fermionic states are connected to four of the eight $\mathbb{Z}_2$ bosonic spin liquid states. Interestingly, the bosonic spin liquid state (which is called $(1,0,0)$ in the main text) closely related to the $(q,0)$ spiral order corresponds to a fermionic spin liquid state (which is called 4b in the main text), while the bosonic state closely related to the $(\pi,\pi)$ spin density wave order does not have a fermionic counterpart. Moreover, all of the root $U(1)$ spin liquid states (where the pairing amplitudes vanish) obtained from the four $\mathbb{Z}_2$ fermionic spin liquid states mentioned above possess Fermi surfaces.

Combining all the information and considering the consistency between different descriptions, we may conclude that the magnetic order below $1 \, \mathrm{K}$ in Volborthite is most likely to be the $(q,0)$ spiral order and the most promising spin liquid state that explains the thermal Hall effect above $1 \, \mathrm{K}$ is the $U(1)$ fermionic spin liquid state with a Fermi surface, which is related to the $(q,0)$ spiral order via the mapping between fermionic and bosonic spin liquid states. The direct confirmation of this conclusion would come from future neutron scattering studies of the excitation spectrum below and above $1 \, \mathrm{K}$.

In the current work, we do not consider the relative energetics of candidate fermionic spin liquid phases, which would require careful examination of projected variational wavefunctions. This would be an excellent topic of future study.

\begin{acknowledgments}
We thank Kyusung Hwang for illuminating discussions. This work was supported by the NSERC of Canada and the Center for Quantum Materials at the University of Toronto. Some of the computations were performed on the GPC supercomputer at the SciNet HPC Consortium.\cite{1742-6596-256-1-012026} SciNet is funded by: the Canada Foundation for Innovation under the auspices of Compute Canada; the Government of Ontario; Ontario Research Fund - Research Excellence; and the University of Toronto.
\end{acknowledgments}

\appendix
\section{\label{spacegroupalgebra} Space Group and Algebraic Identities of Non-symmorphic Kagome Lattice}
We list below the action of spatial symmetries, $T_x$, $T_y$, $C_2$ and $h$, on a site $(x,y,s)$ in the non-symmorphic Kagome lattice.
\begingroup
\allowdisplaybreaks
\begin{align*}
T_x: \: &(x,y,s) \longrightarrow \left(x+1,y,s\right) . \\ \\
T_y: \: &(x,y,s) \longrightarrow \left(x,y+1,s\right) . \\ \\
C_2: \: & (x,y,1) \longrightarrow (-x,-y,5) , \\
& (x,y,2) \longrightarrow (-x,-y,4) , \\
& (x,y,4) \longrightarrow (-x,-y,2) , \\
& (x,y,5) \longrightarrow (-x,-y,1) , \\
& (x,y,3) \longrightarrow (-x+1,-y,3) , \\
& (x,y,6) \longrightarrow (-x,-y+1,6) . \\ \\
h: \: & (x,y,1) \longrightarrow (x,-y-1,2) , \\
& (x,y,2) \longrightarrow (x+1,-y-1,1) , \\
& (x,y,3) \longrightarrow (x,-y,6) , \\
& (x,y,6) \longrightarrow (x+1,-y,3) , \\
& (x,y,4) \longrightarrow (x,-y,5) , \\
& (x,y,5) \longrightarrow (x+1,-y,4) .
\end{align*}
\endgroup
Inspecting how $T_x$, $T_y$, $C_2$ and $h$ commute with each other leads to the algebraic identities
\begingroup
\allowdisplaybreaks
\begin{align}
T_x^{-1} T_y^{-1} T_x T_y &= I , \label{algebraicrelation1} \\
C_2^2 &= I , \label{algebraicrelation2} \\
C_2 T_x C_2^{-1} T_x &= I , \label{algebraicrelation3} \\
C_2 T_y C_2^{-1} T_y &= I , \label{algebraicrelation4} \\
T_x^{-1} h^2 &= I , \label{algebraicrelation5} \\
h^{-1} T_x^{-1} h T_x &= I , \label{algebraicrelation6} \\
h^{-1} T_y h T_y &= I , \label{algebraicrelation7} \\
T_x T_y h^{-1} C_2 h C_2 &= I . \label{algebraicrelation8}
\end{align}
\endgroup
In addition, time reversal commutes with all the space group elements,
\begin{align}
\mathcal{T}^2 &= I , \label{algebraicrelation9} \\
X^{-1} \mathcal{T}^{-1} X \mathcal{T} &= I,\, X \in \lbrace T_x, T_y, C_2, h \rbrace . \label{algebraicrelation10}
\end{align}
\eqref{algebraicrelation1} $-$ \eqref{algebraicrelation10} constrain the possible form of gauge matrices $G_X, X=T_x,T_y,C_2,h,\mathcal{T}$ such that $G_XX \in \mathrm{PSG}$.

\section{\label{fermionicpsgderive} Solution to Fermionic PSG}
The algebraic identities \eqref{algebraicrelation1} $-$ \eqref{algebraicrelation10} lead to a set of consistent conditions among the gauge matrices,
\begingroup
\allowdisplaybreaks
\begin{align}
& G_{T_x}^\dagger(T_y^{-1}(i)) G_{T_y}^\dagger(i) G_{T_x}(i) G_{T_y}(T_x^{-1}(i)) = \eta_{12} \tau^0 , \label{algebraicconstraint1} \\
& G_{C_2}(C_2(i)) G_{C_2}(i) = \eta_{C_2} \tau^0 , \label{algebraicconstraint2} \\
\begin{split}
& G_{C_2}^\dagger(T_xC_2(i)) G_{T_x}(T_xC_2(i)) \\
& \qquad G_{C_2}(C_2(i)) G_{T_x}(i) = \eta_{C_2T_x} \tau^0 , \label{algebraicconstraint3}
\end{split} \\
\begin{split}
& G_{C_2}^\dagger(T_yC_2(i)) G_{T_y}(T_yC_2(i)) \\
& \qquad G_{C_2}(C_2(i)) G_{T_y}(i) = \eta_{C_2T_y} \tau^0 , \label{algebraicconstraint4}
\end{split} \\
& G_{T_x}^\dagger(i) G_h(i) G_h(h^{-1}(i)) = \eta_h \tau^0 , \label{algebraicconstraint5} \\
& G_h^\dagger(T_x^{-1}(i)) G_{T_x}^\dagger(i) G_h(i) G_{T_x}(h^{-1}(i)) = \eta_{hT_x} \tau^0 , \label{algebraicconstraint6} \\
& G_h^\dagger(T_yh(i)) G_{T_y}(T_yh(i)) G_h(h(i)) G_{T_y}(i) = \eta_{hT_y} \tau^0 , \label{algebraicconstraint7} \\
\begin{split}
& G_{T_x}(T_xT_yh^{-1}C_2h(i)) G_{T_y}(T_yh^{-1}C_2h(i)) G_h^\dagger(C_2h(i)) \\
& \qquad G_{C_2}(C_2h(i)) G_h(h(i)) G_{C_2}(i) = \eta_{hC_2} \tau^0 , \label{algebraicconstraint8}
\end{split} \\
& \left[G_\mathcal{T}(i)\right]^2 = \eta_\mathcal{T} \tau^0 , \label{algebraicconstraint9} \\
& G^\dagger_{T_x}(i) G^\dagger_\mathcal{T}(i) G_{T_x}(i) G_\mathcal{T}(T_x^{-1}(i)) = \eta_{T_x\mathcal{T}} \tau^0 , \label{algebraicconstraint10} \\
& G^\dagger_{T_y}(i) G^\dagger_\mathcal{T}(i) G_{T_y}(i) G_\mathcal{T}(T_y^{-1}(i)) = \eta_{T_y\mathcal{T}} \tau^0 , \label{algebraicconstraint11} \\
& G^\dagger_{C_2}(i) G^\dagger_\mathcal{T}(i) G_{C_2}(i) G_\mathcal{T}(C_2^{-1}(i)) = \eta_{C_2\mathcal{T}} \tau^0 , \label{algebraicconstraint12} \\
& G^\dagger_h(i) G^\dagger_\mathcal{T}(i) G_h(i) G_\mathcal{T}(h^{-1}(i)) = \eta_{h\mathcal{T}} \tau^0 , \label{algebraicconstraint13}
\end{align}
\endgroup
where the various $\eta_X=\pm 1$ are $\mathbb{Z}_2$ variables. Using the standard arguments,\cite{PhysRevB.65.165113,PhysRevB.83.224413} we can fix $G_{T_x}(x,y,s)=\eta_{12}^y \tau^0$ and $G_{T_y}(x,y,s)=\tau^0$. Furthermore, we can exploit the global $\mathbb{Z}_2$ gauge such that $G_{T_x} \longrightarrow \eta_h G_{T_x}$ and 
$G_{T_y} \longrightarrow \eta_{hT_y} G_{T_y}$ to fix $\eta_h = \eta_{hC_2} = +1$.

\eqref{algebraicconstraint3} and \eqref{algebraicconstraint4} leads to
\begin{equation} \label{c2recursion}
G_{C_2}(x,y,s) = \eta_{C_2T_x}^x \eta_{C_2T_y}^y \eta_{12}^{\delta_{s,6} x} g_{C_2}(s) ,
\end{equation}
where $g_X(s) \equiv G_X(0,0,s)$. Plugging \eqref{c2recursion} into \eqref{algebraicconstraint2}, we find
\begin{align}
g_{C_2} (C_2(s)) g_{C_2}(s) &= \eta_{C_2} \tau^0,\, s=1,2,4,5, \label{c2square1245} \\
\left[g_{C_2}(3)\right]^2 &= \eta_{C_2} \eta_{C_2T_x} \tau^0 , \label{c2square3} \\
\left[g_{C_2}(6)\right]^2 &= \eta_{C_2} \eta_{C_2T_y} \tau^0 , \label{c2square6}
\end{align}
\eqref{algebraicconstraint6} and \eqref{algebraicconstraint7} leads to
\begin{equation} \label{hrecursion}
G_h(x,y,s) = \eta_{hT_x}^x \eta_{hT_y}^y \eta_{12}^{(\delta_{s,1} +  \delta_{s,2})x} g_h(s) ,
\end{equation}
Applying \eqref{algebraicconstraint5} to $i=(x,y,4)$ and $(x,y,5)$ yields
\begin{align}
\eta_{12}^y \eta_{hT_x} g_h(4) g_h(5) &= \tau^0 , \label{hsquare4} \\
\eta_{12}^y g_h(4) g_h(5) &= \tau^0 , \label{hsquare5}
\end{align}
which implies $\eta_{hT_x}=+1$. Moreover, we must have $\eta_{12}=+1$, since R.H.S. of \eqref{hsquare4} or \eqref{hsquare5} is coordinate independent. Applying \eqref{algebraicconstraint5} to sites with different sublattice index, we find
\begin{align}
g_h(1) g_h(2) &= \eta_{hT_y} \tau^0 , \label{hsquare12} \\
g_h(3) g_h(6) &= \tau^0 , \label{hsquare36} \\
g_h(4) g_h(5) &= \tau^0 . \label{hsquare45}
\end{align}
Applying \eqref{algebraicconstraint8} to $i=(x,y,s)$ for all $s$, using \eqref{c2square1245} to eliminate $g_{C_2}(1) = \eta_{C_2} g_{C_2}^\dagger(5)$ and $g_{C_2}(2) = \eta_{C_2} g_{C_2}^\dagger(4)$, using \eqref{hsquare12}, \eqref{hsquare36} and \eqref{hsquare45} to eliminate $g_h(1)=\eta_{hT_y}g_h^\dagger(2)$, $g_h(3)=g_h^\dagger(6)$ and $g_h(5)=g_h^\dagger(4)$, we find six relations
\begin{align}
g_h^\dagger(4) g_{C_2}(4) g_h(2) g_{C_2}^\dagger(5) &= \eta_{C_2} \eta_{C_2T_y} \tau^0 , \label{hc25} \\
g_h(4) g_{C_2}(5) g_h^\dagger(2) g_{C_2}^\dagger(4) &= \eta_{C_2} \eta_{C_2T_x} \eta_{C_2T_y} \eta_{hT_y} \tau^0 , \label{hc24} \\
g_h^\dagger(6) g_{C_2}(6) g_h(6) g_{C_2}(3) &= \eta_{C_2T_y} \eta_{hT_y} \tau^0 , \label{hc23} \\
g_h(2) g_{C_2}^\dagger(5) g_h^\dagger(4) g_{C_2}(4) &= \eta_{C_2} \eta_{hT_y} \tau^0 , \label{hc22} \\
g_h^\dagger(2) g_{C_2}^\dagger(4) g_h(4) g_{C_2}(5) &= \eta_{C_2} \eta_{C_2T_x} \tau^0 , \label{hc21} \\
g_h(6) g_{C_2}(3) g_h^\dagger(6) g_{C_2}(6) &= \tau^0 . \label{hc26}
\end{align}
Rearranging terms in \eqref{hc23} and \eqref{hc26}, we find that they are equal, which implies $\eta_{hT_y} = \eta_{C_2T_y}$. Similarly, \eqref{hc25} and \eqref{hc24} are related by hermitian conjugation, which implies $\eta_{C_2T_x} = \eta_{C_2T_y}$. \eqref{hc22} and \eqref{hc21} then provide no new information. In summary, there are only four relevant conditions without involving time reversal symmetry
\begin{align}
g_h(4) g_{C_2}(5) g_h^\dagger(2) g_{C_2}^\dagger(4) &= \eta_{C_2} \eta_{C_2T_y} \tau^0 , \label{hc24summary} \\
g_h(6) g_{C_2}(3) g_h^\dagger(6) g_{C_2}^\dagger(6) &= \tau^0 , \label{hc26summary} \\
\left[g_{C_2}(3)\right]^2 &= \eta_{C_2} \eta_{C_2T_y} \tau^0 , \label{c2square3summary} \\
\left[g_{C_2}(6)\right]^2 &= \eta_{C_2} \eta_{C_2T_y} \tau^0 . \label{c2square6summary}
\end{align}

Next, we consider time reversal symmetry. \eqref{algebraicconstraint10} and \eqref{algebraicconstraint11} leads to
\begin{equation} \label{trecursion}
G_\mathcal{T}(x,y,s) = \eta_{T_xT}^x \eta_{T_yT}^y g_\mathcal{T}(s)
\end{equation}
With \eqref{trecursion}, applying \eqref{algebraicconstraint13} on $i=(x,y,s)$ for $s=3$ and $6$ yields
\begin{align}
g_h^\dagger(3) g_\mathcal{T}^\dagger(3) g_h(3) g_\mathcal{T}(6) &= \eta_{T_x\mathcal{T}} \eta_{h\mathcal{T}} \tau^0 , \label{ht3} \\
g_h^\dagger(6) g_\mathcal{T}^\dagger(6) g_h(6) g_\mathcal{T}(3) &= \eta_{h\mathcal{T}} \tau^0 . \label{ht6}
\end{align}
Through \eqref{hsquare36}, we see that \eqref{ht3} and \eqref{ht6} are related by hermitian conjugation, which implies $\eta_{T_x\mathcal{T}} = +1$.

Qualitatively different solutions exist for $\eta_\mathcal{T} = \pm 1$. For $\eta_\mathcal{T} = +1$, $G_\mathcal{T}(i) = \pm \tau^0$ by \eqref{algebraicconstraint9}. From \eqref{timereversalansatzfermion}, the mean field ansatzes satisfy
\begin{equation*}
- u_{ij}^a = G_\mathcal{T}(i) u_{ij}^a G_\mathcal{T}^\dagger(j),\, a=0,x,y,z.
\end{equation*}
For any three sites $i,j,k$ which form a triangle on the non-symmorphic Kagome lattice, any choice of $G_\mathcal{T}(i),G_\mathcal{T}(j),G_\mathcal{T}(k)$ that satisfies $G_\mathcal{T}(\mathbf{r}) = \pm \tau^0$ will render at least one side of the triangle having $u_{ij}^a=0$. This is not a relevant physical solution since it changes the lattice structure, and we will simply ignore it and specialize to the case $\eta_\mathcal{T} = -1$. \eqref{algebraicconstraint9} then implies $G_\mathcal{T}(i) = i \mathbf{a}_i \cdot \bm{\tau}$, where $\mathbf{a}_i$ is a real three-component vector of unit length and $\bm{\tau}$ is the vector of Pauli matrices. We can perform a sublattice dependent gauge transformation $W_s \in SU(2)$ such that $g_\mathcal{T}(s) \longrightarrow W_s g_\mathcal{T}(s) W_s^\dagger = i \tau^2$, without affecting previous gauge fixing.

We can further show that $\eta_{T_y\mathcal{T}}=+1$. Applying \eqref{algebraicconstraint12} to $i=(x,y,s)$ for $s=3$ and $6$ leads to
\begin{align}
g^\dagger_{C_2}(3) g^\dagger_{\mathcal{T}}(3) g_{C_2}(3) g_\mathcal{T}(3) &= \eta_{C_2\mathcal{T}} \tau^0 \label{c2t3} \\ 
g^\dagger_{C_2}(6) g^\dagger_{\mathcal{T}}(6) g_{C_2}(6) g_\mathcal{T}(6) &= \eta_{T_y\mathcal{T}} \eta_{C_2\mathcal{T}} \tau^0 \label{c2t6}
\end{align}
If $\eta_{T_y\mathcal{T}}=-1$, then $g_{C_2}(3)$ and $g_{C_2}(6)$ must have the forms $e^{i \theta \tau^2}$ and $i e^{i \phi \tau^2} \tau^3$, such that \eqref{hc26summary} cannot be satisfied. Therefore, we must have $\eta_{T_y\mathcal{T}}=+1$, which implies $G_\mathcal{T}(i) = i\tau^2$ everywhere.

Now we proceed to solve the equations \eqref{hc24summary} $-$ \eqref{c2square6summary}, plus
\begin{align}
g_{C_2}^\dagger(s) \tau^2 g_{C_2}(s) \tau^2 &= \eta_{C_2 \mathcal{T}} \tau^0 , \label{c2tsummary} \\
g_h^\dagger(s) \tau^2 g_h(s) \tau^2 &= \eta_{h \mathcal{T}} \tau^0 , \label{htsummary}
\end{align}
on a case by case basis. First, note that we have the freedom to perform a sublattice dependent gauge transformation of the form $W_s = e^{i \theta_s \tau^2}$, without affecting previous gauge fixing. The gauge matrices transform as $W_s: g_X(s) \longrightarrow W_s g_X(s) W_{X^{-1}(s)}$ where $X=C_2,h$.

\textbf{A.} $\eta_{C_2}\eta_{C_2T_y} = +1$. \eqref{hc26summary}, \eqref{c2square3summary} and \eqref{c2square6summary} give $g_{C_2}(3) = g_{C_2}(6) = \pm \tau^0$, which further implies $\eta_{C_2 \mathcal{T}}=+1$ by \eqref{c2tsummary}. Therefore, $g_{C_2}(5)=e^{i \theta_5 \tau^2}$ and $g_{C_2}(4)=e^{i \theta_4 \tau^2}$. We perform gauge transformations $W_5 = \pm e^{-i \theta_5 \tau^2}$ and $W_4 = \pm e^{-i \theta_4 \tau^2}$ to fix $g_{C_2}(s) = \pm \tau^0$ for $s=3,4,5,6$, which can be further fixed to $\tau^0$ by a global $\mathbb{Z}_2$ gauge. Moreover, \eqref{hc24summary} gives $g_h(4) = g_h(2)$.

\textbf{i.} $\eta_{h \mathcal{T}}=+1$. \eqref{htsummary} gives $g_h(4) = g_h(2) = e^{i \phi_2 \tau^2}$ and $g_h(6) = e^{i \phi_6 \tau^2}$. Gauge transformations $W_4 = W_2 = e^{-i \phi_2 \tau^2}$ and $W_6 = e^{-i \phi_6 \tau^2}$ fix $g_h(s) = \tau^0$ for $s=2,4,6$. \textbf{(1)}

\textbf{ii.} $\eta_{h \mathcal{T}}=-1$. \eqref{htsummary} gives $g_h(4) = g_h(2) = i e^{i \phi_2 \tau^2} \tau^3$ and $g_h(6) = i e^{i \phi_6 \tau^2} \tau^3$. Gauge transformations $W_4 = W_2 = e^{-i \phi_2 \tau^2}$ and $W_6 = e^{-i \phi_6 \tau^2}$ fix $g_h(s)=i \tau^3$ for $s=2,4,6$. \textbf{(2)}

\textbf{B.} $\eta_{C_2} \eta_{C_2T_y} = -1$.

\textbf{a.} $\eta_{C_2\mathcal{T}} = +1$. \eqref{c2square3summary}, \eqref{c2square6summary} and \eqref{c2tsummary} require that $g_{C_2}(3) = \pm i \tau^2$ and $g_{C_2}(6) = \pm i \tau^2$. \eqref{c2tsummary} gives $g_{C_2}(5)=e^{i \theta_5 \tau^2}$ and $g_{C_2}(4)=e^{i \theta_4 \tau^2}$. Say $g_{C_2}(3) = \pm i \tau^2$, gauge transformations $W_5 = \pm e^{-i \theta_5 \tau^2}$ and $W_4 = \pm e^{-i \theta_4 \tau^2}$ fix $g_{C_2}(5) = g_{C_2}(4) = \pm \tau^0$.

\textbf{i.} $\eta_{h\mathcal{T}} = +1$. \eqref{hc24summary} and \eqref{htsummary} give $g_h(4) = -g_h(2) = e^{i \phi_2 \tau^2}$ and $g_h(6) = e^{i \phi_6 \tau^2}$. Gauge transformations $W_4 = W_2 = e^{-i \phi_2 \tau^2}$ and $W_6 = e^{-i \phi_6 \tau^2}$ fix $g_h(4) = - g_h(2) = g_h(6) = \tau^0$. \eqref{hc26summary} then forces $g_{C_2}(3) = - g_{C_2}(6)$. We fix $g_{C_2}(3) = - g_{C_2}(6)= i \tau^2$ and $g_{C_2}(5) = g_{C_2}(4) = \tau^0$ by and a global $\mathbb{Z}_2$ gauge. \textbf{(3)}

\textbf{ii.} $\eta_{h\mathcal{T}} = -1$. \eqref{hc24summary} and \eqref{htsummary} give $g_h(4) = -g_h(2) = i e^{i \phi_2 \tau^2} \tau^3$ and $g_h(6) = i e^{i \phi_6 \tau^2} \tau^3$. Gauge transformations $W_4 = W_2 = e^{-i \phi_2 \tau^2}$ and $W_6 = e^{-i \phi_6 \tau^2}$ fix $g_h(4) = - g_h(2) = g_h(6) = i \tau^3$. \eqref{hc26summary} then forces $g_{C_2}(3) = g_{C_2}(6)$. We fix $g_{C_2}(3) = g_{C_2}(6) = i \tau^2$ and $g_{C_2}(5) = g_{C_2}(4) = \tau^0$ by a global $\mathbb{Z}_2$ gauge. \textbf{(4)}

\textbf{b.} $\eta_{C_2 \mathcal{T}} = -1$. \eqref{c2tsummary} requires that $g_{C_2}(s) = i e^{i \phi_s \tau^2} \tau^3$. Gauge transformations $W_3 = e^{-i \phi_3 \tau^2/2}$, $W_6 = e^{-i \phi_6 \tau^2/2}$, $W_5 = e^{-i \phi_5 \tau^2}$ and $W_4 = e^{-i \phi_4 \tau^2}$ fix $g_{C_2}(s) = i \tau^3$ for $s=3,4,5,6$.

\textbf{i.} $\eta_{h\mathcal{T}} = +1$. \eqref{hc24summary}, \eqref{hc26summary} and \eqref{htsummary} give $g_h(4) = - g_h^\dagger(2) = e^{i \phi_2 \tau^2}$ and $g_h(6) = \pm i \tau^2$. Gauge transformation $W_4 = W_2^\dagger = \pm e^{-i \phi_2 \tau^2}$ fixes $g_h(4) = -g_h(2) = \pm \tau^0$. We can further fix $g_h(4) = -g_h(2) = \tau^0$ and $g_h(6) = i \tau^2$ by a global $\mathbb{Z}_2$ gauge. \textbf{(5)}

\textbf{ii.} $\eta_{h\mathcal{T}} = -1$. \eqref{hc24summary}, \eqref{hc26summary} and \eqref{htsummary} give $g_h(4) = i e^{i \phi_2 \tau^2} \tau_3$, $g_h(2) = -i e^{-i \phi_2 \tau^2} \tau_3$ and $g_h(6) = \pm i \tau^1$. Gauge transformation $W_4 = W_2^\dagger = \pm e^{-i \phi_2 \tau^2}$ fixes $g_h(4) = -g_h(2) = \pm i \tau^3$. We can further fix $g_h(4) = -g_h(2) = i \tau^3$ and $g_h(6) = i \tau^1$ by a global $\mathbb{Z}_2$ gauge. \textbf{(6)}

We would like to comment on these solutions. First, only $g_{C_2}(s)$ for $s=3,4,5,6$ and $g_h(s)$ for $s=2,4,6$ are explicitly shown here. The remaining gauge matrices can be related through \eqref{c2square1245} and \eqref{hsquare12} $-$ \eqref{hsquare45}. Second, we only consider the overall sign of the product $\eta_{C_2} \eta_{C_2T_y}$ but not the individual components. One of them, say $\eta_{C_2T_y}$, can be $\pm 1$, which doubles the number of solutions to $12$. $\eta_{C_2T_y}$ determines the sign of $g_h(1)$, $g_{C_2}(1)$ and $g_{C_2}(2)$. Finally, we can perform a gauge transformation on the solutions \textbf{(3)}, \textbf{(4)}, \textbf{(5)} and \textbf{(6)}, such that the gauge matrices $g_{C_2}(s)$ and $g_h(s)$ appear more symmetric. For \textbf{(3)} and \textbf{(4)}, we apply the gauge transformations $W_4 = W_5 = i \tau^2$. For \textbf{(5)} and \textbf{(6)}, we apply the gauge transformation $W_6 = -i \tau^2$. The final result is shown in TABLE \ref{fermionicpsgsolution}.

\section{\label{visonpsgderive} Solution to Vison PSG}
To see how vison PSG arises, we start from quantum dimer model (QDM), which depicts spin liquid in a generic lattice as linear combination of singlet product states. QDM can be described by an effective $\mathbb{Z}_2$ gauge theory, which is further mapped to fully frustrated Ising model (FFIM) on the dual lattice, where the notion of vison creation operator becomes apparent. We then construct the dual lattice of non-symmorphic Kagome lattice, which is dubbed the non-symmorphic dice lattice, and solve the vison PSG.

\subsection{\label{qdmtoffim} From Quantum Dimer Model to Fully Frustrated Ising Model}
\begin{figure}
\includegraphics[scale=0.27]{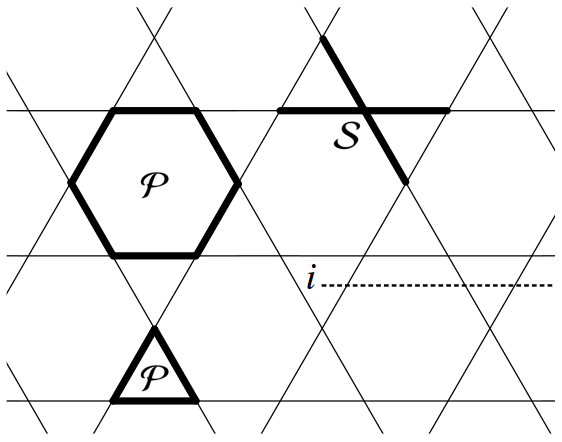}
\caption{\label{plaquettestarvison} Elementary plaquettes $\mathcal{P}$ on Kagome lattice are length-3 triangles and length-6 hexagons. Star $\mathcal{S}$ is defined as the collection of links attached to a site. $i$ labels the sites on dual lattice, which is located at the center of triangular and hexagonal plaquettes. With open boundary condition assumed, the vison creation operator $\sigma_i^z$ involves the product of all links appearing to the right of $i$ and extends to infinity, which is indicated by dashed line.}
\end{figure}

Quantum dimer model (QDM) provides a simple picture of spin liquid in terms of spin singlets (or dimers) for a given lattice. The Hilbert space is spanned by different configurations of dimers, each of which is formed by localized $S=1/2$ moment on two distinct sites of the lattice, such that every site is covered by exactly one dimer (known as hardcore dimer constraint).\cite{PhysRevB.77.134421,PhysRevB.92.205131} In other words, each state in QDM is essentially a linear combination of independent singlet product states on the lattice. The Hamiltonian of QDM has a kinetic term, which changes the configuration of dimers, and a potential term, which counts the interaction between dimers.\cite{PhysRevB.77.134421,PhysRevB.92.205131} QDM can be effectively described by a $\mathbb{Z}_2$ gauge theory through the introduction of Pauli matrices $\bm{\tau}_l$ defined on each link $l$ connecting two sites on the lattice, such that $\tau_l^x=-1$ ($\tau_l^x=+1$) when $l$ is occupied (unoccupied) by a dimer, while $\tau_l^z$ changes the state of $l$. This leads to the Hamiltonian \cite{PhysRevB.77.134421}
\begin{equation} \label{z2gaugetheoryhamiltonian}
H = - \sum_l J_l \tau_l^x - \sum_{\mathcal{P}} \Gamma_\mathcal{P} \prod_{l \in \mathcal{P}} \tau_l^z ,
\end{equation}
where $\mathcal{P}$ denotes an elementary plaquette, which is either a triangle or a hexagon on Kagome lattice (FIG. \ref{plaquettestarvison}). The first term in \eqref{z2gaugetheoryhamiltonian} corresponds to the potential term while the second term corresponds to the kinetic term. In addition, the hardcore dimer constraint requires \cite{PhysRevB.77.134421}
\begin{equation} \label{hardcoredimerconstraint}
\prod_{l \in \mathcal{S}} \tau_l^x = -1 ,
\end{equation}
where $\mathcal{S}$ denotes a `star', the collection of links attached to a given site on the lattice (FIG. \ref{plaquettestarvison}). Note that the Hamiltonian \eqref{z2gaugetheoryhamiltonian} respects the hardcore dimer constraint \cite{PhysRevB.77.134421,PhysRevB.92.205131}
\begin{equation*}
\left[ H, \prod_{l \in \mathcal{S}} \tau_l^x \right] = 0 .
\end{equation*}

$\mathbb{Z}_2$ gauge theory on the \textit{direct} lattice as defined above can be further mapped to fully frustrated Ising model (FFIM) on the \textit{dual} lattice, which is dice lattice in the case of Kagome lattice. This is done by introducing the operators \cite{PhysRevB.77.134421}
\begin{align}
\sigma_i^x &= \prod_{l \in \mathcal{P}(i)} \tau_l^z , \label{sigmaxdefinition} \\
\sigma_i^z &= \prod_{l > i} \tau_l^x , \label{sigmazdefinition}
\end{align}
where $i$ label the sites on dual lattice, which are the center of plaquettes $\mathcal{P}$ on direct lattice. $\bm{\sigma}$ obeys the same Pauli matrices algebra as $\bm{\tau}$.\cite{PhysRevB.92.205131} $\sigma_i^z$, which involves the product of all links appearing to the right of $i$ (FIG. \ref{plaquettestarvison}), is nonlocal in the direct lattice but local in the dual lattice. In $\mathbb{Z}_2$ gauge theory, the flux of an elementary plaquette on the direct lattice is given by $\sigma_i^x$. If $i$ is the center of $\mathcal{P}$, then $\sigma_i^z$ changes the flux of $\mathcal{P}$ by $\pm 1$, and we say that a vison is created. Therefore, $\sigma_i^z$ is understood as the vison creation operator. The definition \eqref{sigmazdefinition} and the constraint \eqref{hardcoredimerconstraint} imply \cite{PhysRevB.77.134421}
\begin{equation}
\sigma_i^z \sigma_j^z = \lambda_{ij} \tau_l^x
\end{equation}
for two neighboring sites $i$ and $j$ separated by link $l$, where $\lambda_{ij}=\pm 1$ is chosen such that the Gauss law constraint, the equivalent of hardcore dimer constraint in FFIM, is satisfied. Gauss law constraint states that the product of $\lambda_{ij}$ around an elementary plaquette on the dual lattice has to be $-1$ (to see this, write $\tau_l^x= \lambda_{ij} \sigma_i^z \sigma_j^z$ and apply \eqref{hardcoredimerconstraint}). \eqref{z2gaugetheoryhamiltonian} can then be written as
\begin{equation} \label{ffimhamiltonian}
H = - \sum_{\langle ij \rangle} \lambda_{ij} J_{ij} \sigma_i^z \sigma_j^z - \sum_i \Gamma_i \sigma_i^x ,
\end{equation}
which is the Hamiltonian of FFIM on the dual lattice. Following Ref.~\onlinecite{PhysRevB.84.094419}, we solve \eqref{ffimhamiltonian} for the vison dispersion with soft spin approximation, in which $\sigma_i^z = \pm 1$ is replaced by a continuous variable $\phi_i \in \mathbb{R}$. We also neglect the second term in \eqref{ffimhamiltonian} as our symmetry consideration are restricted to time independent and static configurations. Therefore, \eqref{ffimhamiltonian} becomes
\begin{equation} \label{softspinhamiltonian}
H = - \sum_{\langle ij \rangle} \lambda_{ij} J_{ij} \phi_i \phi_j .
\end{equation}

\subsection{\label{nonsymmorphicdice} Non-Symmorphic Dice Lattice}
\begin{figure}
\begin{tabular}{>{\centering\arraybackslash} m{4.0 cm} >{\centering\arraybackslash} m{3.5 cm}}
\includegraphics[scale=0.28]{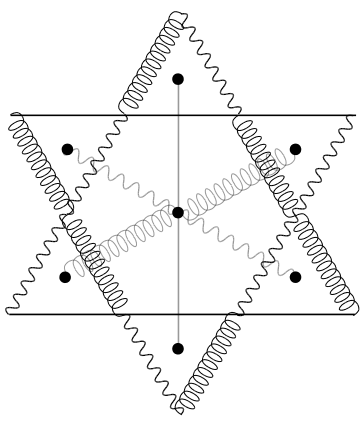} & \includegraphics[scale=0.3]{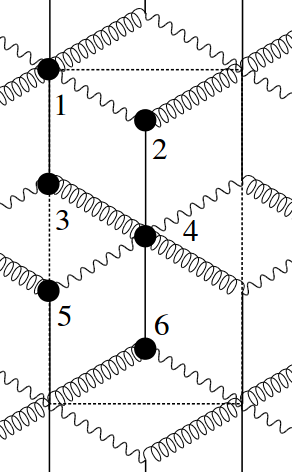} \\
(a) & (b)
\end{tabular}
\caption{\label{dualdice} The dual lattice of non-symmorphic Kagome lattice is constructed by (a) connecting the center of triangles and hexagons across the links, which results in (b) non-symmorphic dice lattice with three inequivalent links, six sites per unit cell (indicated by dashed line) and the same space group.}
\end{figure}

Non-symmorphic Kagome lattice can be constructed from isotropic Kagome lattice by replacing the nearest neighbour bonds of the latter with three different bonds. Since the mapping from QDM to FFIM need not preserve the microscopic interactions, we can forget about the spin model suggested by DFT calculation for the moment, and view the non-symmorphic Kagome lattice as consisting of three generically inequivalent links. Its dual lattice, the non-symmorphic dice lattice, is constructed by connecting the center of triangular and hexagonal plaquettes across these links, as shown in FIG. \ref{dualdice}a. As a result, non-symmorphic dice lattice also has three generically inequivalent links, six sites per unit cell (FIG. \ref{dualdice}b), and the same space group as non-symmorphic Kagome lattice.

\begin{figure}
\begin{tabular}{>{\centering\arraybackslash} m{4.6 cm} >{\centering\arraybackslash} m{3.4 cm}}
\includegraphics[scale=0.27]{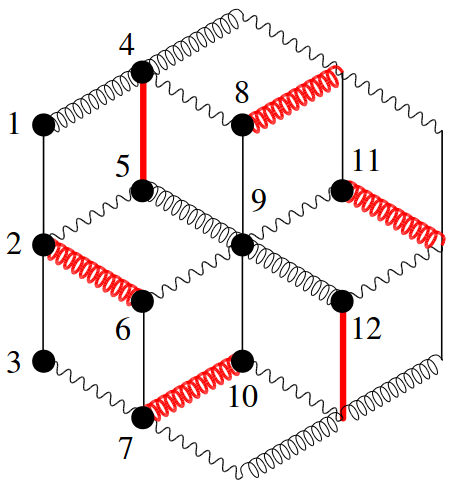} & \includegraphics[scale=0.18]{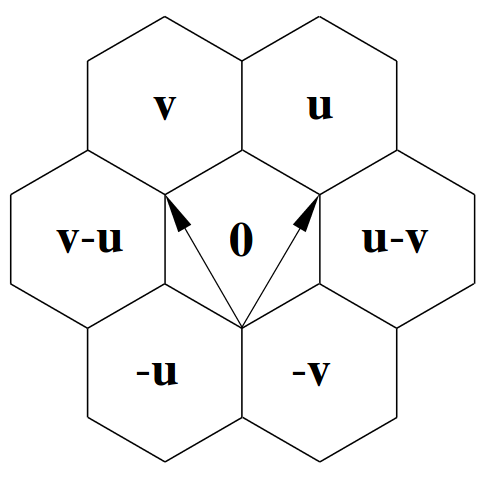} \\
\ \ \ \ (a) & (b)
\end{tabular}
\caption{\label{gaugehoneycomb} (a) The gauge we choose on the non-symmorphic Dice lattice in accordance with Gauss law constraint. Red thick (black thin) links have $\lambda_{ij}=-1$ ($\lambda_{ij}=+1$), such that the product of $\lambda_{ij}$ around each rhombus is $-1$. This defines a hexagonal unit cell with twelve sites. (b) The effective honeycomb lattice and primitive vectors $\mathbf{u}$ and $\mathbf{v}$.}
\end{figure}

Elementary plaquettes on the non-symmorphic dice lattice are the different rhombi (FIG. \ref{dualdice}b). We must now choose a combination of $\lambda_{ij}=\pm 1$ such that Gauss law constraint is satisfied. It turns out that the gauge introduced in Ref.~\onlinecite{PhysRevB.84.094419} is a convenient choice, which we adapt to our system as depicted in FIG. \ref{gaugehoneycomb}a. This particular gauge choice enlarges the unit cell such that the original dice lattice effectively becomes a honeycomb lattice with twelve sites per unit cell. Primitive vectors $\mathbf{u}$ and $\mathbf{v}$ duplicate the unit cell in two independent directions (FIG. \ref{gaugehoneycomb}b), so that the coordinates of any unit cell is $\mathbf{R} = m\mathbf{u} + n\mathbf{v}$ where $m,n \in \mathbb{Z}$.

\begin{figure}
\begin{tabular}{c}
\includegraphics[scale=0.27]{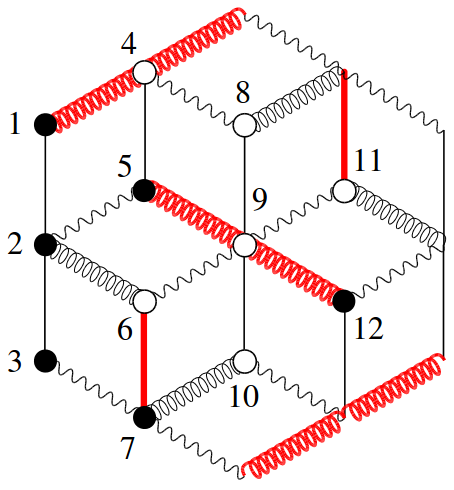} \\
\ \ \ \ (a) \\ \\
\includegraphics[scale=0.27]{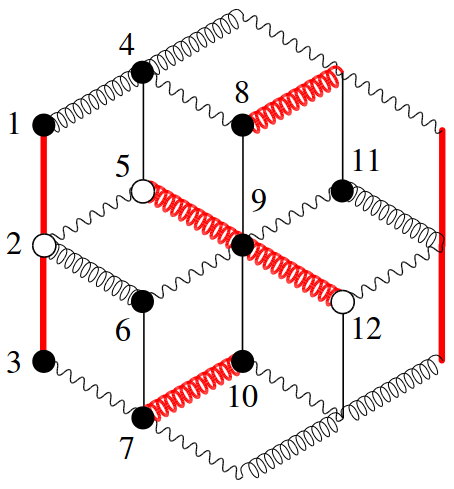} \\
\ \ \ \ (b)
\end{tabular}
\caption{\label{gaugechange} Unit cell of the effective honeycomb lattice under (a) translations $T_x$ and $T_y$ and (b) glide $h$. Red thick (black thin) links have $\lambda_{ij}=-1$ ($\lambda_{ij}=+1$). To restore the original gauge choice (FIG. \ref{gaugehoneycomb}a), we have to multiply $-1$ to the vison field components that corresponds to sites indicated by empty circles.}
\end{figure}

Consider the action of symmetry transformations of non-symmorphic dice lattice on the effective honeycomb lattice. FIG. \ref{gaugechange}a and b shows how our original gauge choice is modified by translations and glide respectively, while it is invariant under $\pi$-rotation and time reversal. To restore the original gauge choice, we have to multiply certain vison field components $\phi_i$ by the factor $-1$ as indicated in FIG. \ref{gaugechange}a and b, which corresponds to a $\mathbb{Z}_2$ gauge transformation. Such a combination of symmetry and gauge transformation that leaves the Hamiltonian \eqref{softspinhamiltonian} invariant defines the vison PSG.

Next, we extract the matrix representation of vison PSG in the order parameter space that describes vison condensation.\cite{PhysRevB.92.205131} The Hamiltonian \eqref{softspinhamiltonian} in momentum space is obtained through Fourier transform,
\begin{equation*}
\begin{aligned}
H &= - \sum_{\left\langle ij \right\rangle} \lambda_{ij} J_{ij} \phi_i \phi_j \\
&= - \sum_{i} \sum_{\mathbf{kk}'} \sum_{\left(a,b,\delta_\mathrm{I},\delta_\mathrm{II},\lambda,\mathrm{T}\right)} \lambda J_\mathrm{T} \phi_\mathbf{k}^{a} \frac{e^{i\mathbf{k} \cdot \left(\mathbf{R}_i+\delta_\mathrm{I} \right)}}{\sqrt{N}} \phi_{\mathbf{k}'}^{b} \frac{e^{i\mathbf{k}' \cdot \left(\mathbf{R}_i+\delta_\mathrm{II} \right)}}{\sqrt{N}} \\
&= - \sum_\mathrm{k} \sum_{\left(a,b,\mathbf{d},\lambda,\mathrm{T}\right)} \lambda J_\mathrm{T} \phi_\mathbf{k}^{a} \phi_{-\mathbf{k}}^{b} e^{i\mathbf{k}\cdot\mathbf{d}} \\
&\equiv \sum_\mathrm{k} \sum_{ab} \phi_{\mathbf{k}}^{a} \mathrm{D}_\mathbf{k}^{ab} \phi_{-\mathbf{k}}^{b} ,
\end{aligned}
\end{equation*}
where $a,b=1,...,12$ are sublattice indices, $\delta_\mathrm{I}$ and $\delta_\mathrm{II}$ are displacements with respect to coordinate $\mathbf{R}_i$ of unit cell $i$, $\mathbf{d} \equiv \delta_\mathrm{I}-\delta_\mathrm{II}$, $\lambda=\pm 1$ reflects the gauge choice in FIG. \ref{gaugehoneycomb}a, and $\mathrm{T}=\mathrm{I},\mathrm{II},\mathrm{III}$ labels the three inequivalent couplings. Diagonalization of $\mathrm{D}_\mathbf{k}$ yields the vison dispersion. The order parameters $\psi_n$ are defined through the expansion of vison field in terms of the critical modes,\cite{PhysRevB.92.205131}
\begin{equation}
\phi \left(\mathbf{R}\right) = \sum_{n} \psi_n \mathbf{v}_n e^{i \mathbf{q}_n \cdot \mathbf{R}} ,
\end{equation}
where we label the wave vectors at which minimum occurs by $\mathbf{q}_n$, and the corresponding eigenvector by $\mathbf{v}_n$. These eigenvectors are chosen such that they form a complete orthonormal set, $\sum_n \mathbf{v}_n \mathbf{v}_n^\dagger = 1$. The vison PSG can be viewed as matrix operation on the order parameters as follows,\cite{PhysRevB.84.094419,PhysRevB.92.205131}
\begin{equation}
\begin{aligned}[b]
G_X X : \phi^a (\mathbf{R})
\longrightarrow & \sum_{n} \psi_n v_n^{a'} e^{i \mathbf{q}_n \cdot \mathbf{R}'} G_X \left(\mathbf{R}',a'\right) \\
= & \sum_{n} \left( \sum_{m} O_{nm} \psi_m \right) v_n^{a} e^{i \mathbf{q}_n \cdot \mathbf{R}} ,
\end{aligned}
\end{equation}
where $\left(\mathbf{R}',a'\right)= X \left(\mathbf{R},a\right)$. $O_{nm}$ is the matrix representation of vison PSG that we want to determine. With the couplings $J_\mathrm{I}, J_\mathrm{II}, J_\mathrm{III}$ chosen arbitrarily, we find only two critical modes at $\mathbf{q}_1 = \mathbf{q}_2 = \mathbf{0}$. The PSG matrices are then constructed from the corresponding eigenvectors $\mathbf{v}_1$ and $\mathbf{v}_2$,
\begin{align}
T_x &=
\begin{pmatrix}
0 & -1 \\
1 & 0
\end{pmatrix} , \label{visontranslationxpsg} \\
T_y &=
\begin{pmatrix}
0 & -1 \\
1 & 0
\end{pmatrix} , \\
C_2 &=
\begin{pmatrix}
1 & 0 \\
0 & 1
\end{pmatrix} , \\
h &= \frac{1}{\sqrt{2}}
\begin{pmatrix}
-1 & -1 \\
1 & -1
\end{pmatrix} , \\
\mathcal{T} &=
\begin{pmatrix}
1 & 0 \\
0 & 1
\end{pmatrix} , \label{visontimereversalpsg}
\end{align}
which result in the symmetry fractionalizations
\begingroup
\allowdisplaybreaks
\begin{align}
T_x^{-1} T_y^{-1} T_x T_y &= 1 , \label{visonsymmetryfractionalization1} \\
C_2^2 &= 1 \\
C_2 T_x C_2^{-1} T_x &= -1 , \\
C_2 T_y C_2^{-1} T_y &= -1 , \\
T_x^{-1} h^2 &= -1 , \\
h^{-1} T_x^{-1} h T_x &= 1 , \\
h^{-1} T_y h T_y &= -1 , \\
T_x T_y h^{-1} C_2 h C_2 &= -1 , \\
\mathcal{T}^2 &= 1 , \\
T_x^{-1} \mathcal{T}^{-1} T_x \mathcal{T} &= 1 , \\
T_y^{-1} T^{-1} T_y \mathcal{T} &= 1 , \\
C_2^{-1} \mathcal{T}^{-1} C_2 \mathcal{T} &= 1 , \\
h^{-1} \mathcal{T}^{-1} h \mathcal{T} &= 1 . \label{visonsymmetryfractionalization13}
\end{align}
\endgroup
Furthermore, we find that while different sets of couplings $\lbrace J_\mathrm{I}, J_\mathrm{II}, J_\mathrm{III} \rbrace$ may change the matrix representations \eqref{visontranslationxpsg} $-$ \eqref{visontimereversalpsg} of vison PSG, they lead to the same symmetry fractionalizations \eqref{visonsymmetryfractionalization1} $-$ \eqref{visonsymmetryfractionalization13}.

\section{\label{fusionrule} Trivial and Nontrivial Fusion Rule}
We check the fusion rule of the algebraic identities listed in Appendix \ref{spacegroupalgebra}. Let us consider the motion of a fermionic spinon $f$, which is the bound state of a bosonic spinon $b$ and vison $v$ as discussed in Section \ref{bosonfermionmap}, under these algebraic identities. We are going to focus on the relative motion between $b$ and $v$ rather than the graphical derivation introduced in Ref.~\onlinecite{1603.03041,1605.05322}, since the conclusion of the latter seems to depend on how $b$ is placed relative to $v$ initially.

First, notice that there is no relative motion between $b$ and $v$ under translations. Therefore, \eqref{algebraicrelation1} has trivial fusion rule. Next, consider a unitary operator $X$ that squares to identity, for instance $C_2$. We argue that $X^2=I$ must have nontrivial fusion rule as follows.\cite{1403.0575,1605.05322} Consider a state $\left\lvert \Psi \right\rangle = f_i^\dagger f_{X(i)}^\dagger \left\lvert 0 \right\rangle$ with $\left\lvert 0 \right\rangle$ being the ground state. Under the action of $X$,
\begin{equation}
X: f_i^\dagger \longrightarrow X f_i^\dagger X^{-1} \equiv f_{X(i)}^\dagger
\end{equation}
and
\begin{equation}
\begin{aligned}[b]
X: \left\lvert \Psi \right\rangle &\longrightarrow \left(X f_i^\dagger X^{-1}\right) \left(X f_{X(i)}^\dagger X^{-1}\right) \left\lvert 0 \right\rangle \\
&= f_{X(i)} X^2 f_{i} X^{-2} \left\lvert 0 \right\rangle \\
&= e^{i\phi_f} f_{X(i)} f_{i} \left\lvert 0 \right\rangle \\
&= - e^{i\phi_f} \left\lvert \Psi \right\rangle ,
\end{aligned}
\end{equation}
where the minus sign in the last line arises from exchanging two fermions. If we write $f_i^\dagger = b_i^\dagger v_i^\dagger$, then we have
\begin{equation}
X: \left\lvert \Psi \right\rangle \longrightarrow e^{i\phi_b} e^{i\phi_v} \left\lvert \Psi \right\rangle ,
\end{equation}
since bosonic operators commute. This shows that $X^2=I$ must have nontrivial fusion rule $- e^{i\phi_f} = e^{i\phi_b} e^{i\phi_v}$ for a unitary operator $X$. Therefore, \eqref{algebraicrelation2} have nontrivial fusion rule. An equivalent loop traced by \eqref{algebraicrelation3} is $C_2^{-1} T_x C_2 T_x = (C_2^{-1})^2 (C_2 T_x)^2$, where the operator in each bracket squares to identity. Therefore, \eqref{algebraicrelation3} has trivial fusion rule. The same argument applies to \eqref{algebraicrelation4}.

$h$ is reflection followed by half lattice translation. We conjecture that, since reflection alone squares to identity, and all sorts of translation do not cause any relative motion between $b$ and $v$, \eqref{algebraicrelation5} is nontrivial because it involves reflection twice. It also follows that \eqref{algebraicrelation6} and \eqref{algebraicrelation7} are trivial, because reflection is first applied and its inverse subsequently. \eqref{algebraicrelation8} is nontrivial because it involves rotation twice, and reflection and its inverse once.

Since $\mathcal{T}$ is antiunitary, we cannot conclude from previous argument that \eqref{algebraicrelation9} has nontrivial fusion rule. In fact, it has trivial fusion rule, since $\mathcal{T}^2=-1$ for half-integer spins (e.g. $b$ and $f$), while $\mathcal{T}^2=+1$ for integer spin (e.g. $v$).\cite{1403.0575} For $X=T_x,T_y$, \eqref{algebraicrelation10} has trivial fusion rule. For $X=C_2,h$, \eqref{algebraicrelation10} has nontrivial fusion rule, because, considering the equivalent loop $X \mathcal{T} X^{-1} \mathcal{T}^{-1}$, the effect of $\mathcal{T} X^{-1} \mathcal{T}^{-1}$ should be the same as $X$, so the twist factor is equal to that under $X^2$.

\bibliography{draft}
\end{document}